\documentclass[pra,aps,twocolumn,superscriptaddress,showpacs]{revtex4-1}
\usepackage{amsmath}
\usepackage{graphicx}
\usepackage{amssymb}
\usepackage{graphics}
\usepackage{amsthm}

\begin{document}
\title{Randomly poled crystals as a source of photon pairs}

\author{Jan Pe\v{r}ina Jr.}
\author{Ji\v{r}\'{\i} Svozil\'{\i}k}
\affiliation{Joint Laboratory of Optics, Palack\'{y} University
and Institute of Physics of Academy of Sciences of the Czech
Republic, 17. listopadu 50a, 772 07 Olomouc, Czech Republic}

\email{perinaj@prfnw.upol.cz}

\begin{abstract}
Generation of photon pairs from randomly poled nonlinear crystals
is investigated using analytically soluble model and numerical
calculations. Randomly poled crystals are discovered as sources of
entangled ultra broad-band signal and idler fields. Their
photon-pair generation rates scale linearly with the number of
domains. Entanglement times as short as several fs can be reached.
Comparison with chirped periodically-poled structures is given and
reveals close similarity.
\end{abstract}


\maketitle

\section{Introduction}

The first experimentally observed nonlinear optical effect,
second-harmonic generation, was investigated already more than
forty years ago by Franken \cite{Franken1961}. Since that, many
other nonlinear effects have been revealed and understood even at
quantum level. Among them, spontaneous parametric down-conversion
(SPDC) with its production of photon pairs belongs to the most
fascinating. The reason is that two photons comprising a photon
pair generated in one quantum event of this process are mutually
strongly correlated (entangled) as was discovered by Hong, Ou and
Mandel in eighties \cite{Ghosh1986}. They used in their
experiments nonlinear bulk crystals that later become the most
common sources of photon pairs \cite{Rubin1994}. In order to
observe spontaneous parametric down-conversion phase-matching
conditions of the interacting three optical fields have to be
fulfilled. Unfortunately, they cannot be naturally fulfilled in
many highly nonlinear crystals. However, Armstrong
\cite{Armstrong1962} has arrived with the concept of additional
periodic modulation of nonlinear susceptibility that has been
practically developed in periodical poling of nonlinear crystals
\cite{Hum2007}. In this concept, wave vector of the additionally
introduced nonlinear modulation is added to the natural
phase-matching condition and the so-called quasi-phase-matching of
the interacting optical fields is reached this way. Highly
nonlinear materials can be exploited since then. We note that
shortening of a nonlinear medium such that phase-matching
conditions loose their importance is the only alternative way to
periodical poling. This approach has been applied in
photonic-band-gap structures in which optical interference is
crucial for reaching an efficient nonlinear interaction
\cite{PerinaJr2006,Centini2005,Vamivakas2004,PerinaJr2007b,PerinaJr2009,PerinaJr2009a}.

Periodical poling has occurred to be extraordinarily useful. It
has provided not only compensation for the natural phase mismatch.
The ability to tailor the properties of emitted optical fields has
been revealed soon. It is based on using ordered nonlinear domains
with variable lengths (chirped periodical poling). Presence of
domains of different lengths in an ordered structure allows an
efficient nonlinear interaction of fields in a broad spectral
range. For example, signal and idler fields with spectra several
hundreds of nm wide can be generated in chirped LiNbO$_{3}$
crystals. On the other hand, domains with different lengths can
also be ordered randomly. A bit surprisingly, the nonlinear
interaction can be efficient even in this case sometimes referred
as stochastic quasi-phase-matching (SQPM). Similarly as ordered
poled structures the randomly poled structures (RPS) allow
spectrally broadband nonlinear interaction. It is not surprising
that efficiency of random structures is worse compared to ordered
ones. However, they usually put smaller requirements to
polarization properties of the interacting optical fields as well
as orientation of the nonlinear medium
\cite{Baudrier-Raybaut2004}. Also fabrication of RPS is much
easier because high precision is required in production of ordered
periodically-poled structures.

The role of SQPM in 1D has already been addressed for the process
of second-harmonic generation
\cite{Morozov2004,Vidal2006,Aleksandrovsky2008,Centini2006,Kitaeva2007}
and the process of difference-frequency generation
\cite{Pelc2010}. Moreover full domain random structures allowing
SQPM for transversal second-harmonic generation have been studied
in \cite{Fischer2008}.

Here, we focus our attention to the generation of photon pairs in
randomly poled 1D nonlinear crystals \cite{Svozilik2010}. It is
shown that spectral properties of photon pairs and photon-pair
generation rates are comparable in RPS and chirped
periodically-poled structures (CPPS)
\cite{Nasr2008,Harris2007,Svozilik2009,Saleh2009}. This is very
promising for easy production of broadband and efficient sources
of photon pairs. These broadband sources are important, e.g., in
metrology (quantum optical coherence tomography
\cite{Carrasco2004}) or quantum-information processing
\cite{Humble2007,Rohde2005}. We note that broadband photon-pair
sources can also be constructed using zero group-velocity
dispersion conditions \cite{Donnell2007}. However, such conditions
can be met only for certain pump frequencies considering a given
material.

The paper is organized as follows. In Sec.~II, a general theory of
SPDC modified to random structures is presented. Photon-pair
generation rates and intensity spectra are studied both for random
and chirped structures in Sec.~III. Sec.~IV is devoted to temporal
properties of the generated photon pairs. Spatial properties of
photon pairs are addressed in Sec.~V. Temperature behavior of the
quantities characterizing photon pairs is analyzed in Sec.~VI.
Sec.~VII brings the analysis of fabrication errors. The role of
ordering in chirped structures is studied in Sec.~VIII. Finally,
conclusions are drawn in Sec.~IX.

\section{Spontaneous parametric down-conversion in poled nonlinear crystals}

The process of SPDC in a nonlinear crystal can be conveniently described
by the following interaction Hamiltonian $\hat{H}_{{\rm int}}$ \cite{Hong1985,Saleh1991}:
\begin{eqnarray}  
\hat{H}_{{\rm int}}(t) & = & \varepsilon_{0}{\cal B}\int_{-L}^{0}dz\chi^{(2)}(z)\nonumber \\
 &  & \hspace{-15mm}\mbox{}\times E_{p}^{(+)}(z,t)\hat{E}_{s}^{(-)}(z,t)\hat{E}_{i}^{(-)}(z,t)+{\rm H.c.};
\label{1}
\end{eqnarray}
$L$ denotes the crystal length. In Eq.~(\ref{1}), the
positive-frequency part of the pump electric-field amplitude is
denoted as ${\bf E}_{p}^{(+)}$ and ${\bf E}_{s}^{(-)}$ (${\bf
E}_{i}^{(-)}$) stands for the negative-frequency part of the
signal (idler) electric-field amplitude operator. The
$z$-dependent second-order susceptibility tensor $\chi^{(2)}$
describes poling of the nonlinear material. Vacuum permittivity is
denoted as $\varepsilon_{0}$ and ${\cal B}$ means the transverse
area of the optical fields. Symbol ${\rm H.c.}$ replaces the
Hermitian-conjugated term.

Electric-field amplitudes occurring in Eq.~(\ref{1}) can be conveniently
decomposed into harmonic plane waves with frequencies $\omega_{a}$
and wave vectors $k_{a}$:
\begin{eqnarray}  
  \hat{E}_{a}^{(-)}(z,t) & = & \frac{1}{2\pi}\int
   d\omega_{a}\hat{E}_{a}^{(-)}(\omega_{a})\exp(-ik_{a}z+i\omega_ {a}t),\nonumber
  \\
 &  & \hspace{2cm}a=s,i.
\label{2}
\end{eqnarray}
The quantum spectral amplitudes $\hat{E}_{a}^{(-)}(\omega_{a})$ in
Eq.~(\ref{2}) can then be expressed using photon creation
operators $\hat{a}_{a}^{\dagger}(\omega_{a})$:
\begin{equation}  
 \hat{E}_{a}^{(-)}(\omega_{a})=-i\sqrt{\frac{\hbar\omega_{a}}{2\varepsilon_{0}cn_
 { a}(\omega_{a}){\cal
 B}}}\hat{a}_{a}^{\dagger}(\omega_{a});
\label{3}
\end{equation}
 $c$ is speed of light in vacuum, $\hbar$ reduced Planck constant,
and $n_{a}$ stands for index of refraction of field $ a $.

First-order perturbation solution of Schr\"{o}dinger equation with
the initial vacuum state $|{\rm vac}\rangle$ in the signal and
idler fields results in the following two-photon quantum state
$|\psi\rangle$:
\begin{eqnarray}   
 |\psi\rangle = \int d\omega_{s}\int d\omega_{i}{\Phi}(\omega_{s},\omega_{i})
 \hat{a}_{s}^{\dagger}(\omega_{s})\hat{a}_{i}^{\dagger}(\omega_{i})|{\rm
  vac}\rangle.
\label{4}
\end{eqnarray}
The two-photon spectral amplitude $\Phi$ introduced in
Eq.~(\ref{4}) is given as follows:
\begin{eqnarray}   
 {\Phi}(\omega_{s},\omega_{i}) & = &
  g(\omega_{s},\omega_{i})E_{p}^{(+)}(\omega_{s}+\omega_{i})\nonumber\\
 &  & \mbox{}\times
  F(\Delta k(\omega_{s},\omega_{i})),
\label{5}
\end{eqnarray}
where $g$ denotes a coupling constant,
$g(\omega_{s},\omega_{i})=i\sqrt{\omega_{s}\omega_{i}}/[2c\pi\sqrt{n_{s}(\omega_
{s})n_{i}(\omega_{i})}]\chi^{(2)}(0)$, and
$E_{p}^{(+)}(\omega_{p})$ stands for the pump-field amplitude
spectrum. The stochastic function $F$ introduced in Eq.~(\ref{5})
describes SQPM and is derived in the form \cite{Svozilik2010}:
\begin{equation}  
 F(\Delta k)=\int_{-L}^{0}dz\,\frac{\chi^{(2)}(z)}{\chi^{(2)}(0)}\exp(i\Delta
 kz).
\label{6}
\end{equation}
Symbol $\Delta k$ describes the natural phase mismatch for the interacting
fields, $\Delta k=k_{p}-k_{s}-k_{i}$.

In a periodically-poled structure neighbor domains differ in signs
of $\chi^{(2)}$ nonlinearity and function $F$ defined in Eq.~(\ref{6})
can be recast into the form:
\begin{eqnarray}   
 F(\Delta k)=\sum_{n=1}^{N_{L}}(-1)^{n-1}\int_{z_{n-1}}^{z_{n}}dz\exp(i\Delta
 kz).
\label{7}
\end{eqnarray}
Symbol $N_{L}$ denotes the number of domains and $n$-th domain
extends from $z=z_{n-1}$ to $z=z_{n}$. Positions $z_{n}$ of domain
boundaries are random and can be expressed as
$z_{n}=z_{n-1}+l_{0}+\delta l_{n}$ ($n=1,\ldots,N_{L}$,
$z_{0}=-L$) in our model using stochastic Gaussian declinations
$\delta l_{n}$. The basic layer length $l_{0}$ is determined such
that quasi-phase-matching is reached, i.e. $l_{0}=\pi/\Delta
k_{0}$, $\Delta k_{0}\equiv\Delta
k(\omega_{s}^{0},\omega_{i}^{0})$, and $\omega_{a}^{0}$ means
central frequency of field $a$. The random declinations $\delta
l_{n}$ are mutually independent and can be described by the joint
Gaussian probability distribution $P$:
\begin{equation}  
 P(\delta{\bf L})=\frac{1}{(\sqrt{\pi}\sigma)^{N_{L}}}\exp(-\delta{\bf
 L}^{T}{\bf B}\delta{\bf L}).
\label{8}
\end{equation}
Covariance matrix ${\bf B}$ is assumed to be diagonal and its
nonzero elements equal $1/\sigma^{2}$. Stochastic vector
$\delta{\bf L}$ is composed of declinations $\delta l_{n}$; symbol
$^{T}$ stands for transposition. Characteristic function $G$ of
the distribution $P$ in Eq.~(\ref{8}) takes the form:
\begin{equation}   
 G(\delta{\bf K})\equiv\langle\exp(i\delta{\bf K}\cdot\delta{\bf
 L})\rangle_{{\rm av}}=\prod_{j=1}^{N}G(\delta k_{j});
\label{9}
\end{equation}
symbol $ \cdot $ means scalar product. Vector $\delta{\bf K}$ of
parameters of the characteristic function $G$ is composed of
elements $\delta k_{j}$. One-dimensional characteristic function
$G(\delta k)$ in Eq.~(\ref{9}) equals $\exp(-\sigma^{2}\delta
k^{2}/4)$.

In order to obtain analytical formulas, we integrate the expression
for function $F$ in Eq.~(\ref{7}) domain by domain and modify the
contributions of the first and last domains in such a way that the
following simple formula is reached:
\begin{equation}  
 F(\Delta k)=\frac{2i}{\Delta k}\sum_{n=0}^{N_{L}}(-1)^{n}\exp(i\Delta
 kz_{n}).
\label{10}
\end{equation}
As a typical structure contains hundreds of domains, incorrect
inclusion of fields from the first and the last domains leads to
negligible declinations. The formula in Eq.~(\ref{10}) can be
interpreted such that SPDC occurs only in domains with positive
susceptibility $\chi^{(2)}$ at doubled amplitude and domains with
negative susceptibility $\chi^{(2)}$ play only the role of a
'linear' filler. This interpretation elucidates why a pair of
domains having one positive and one negative signs of
susceptibility $\chi^{(2)}$ forms an elementary unit for
understanding properties of photon pairs.

\section{Photon-pair generation rates and intensity spectra}

Photon-pair generation rate as well as intensity spectra can be easily
derived from mean spectral density of the number of generated photon
pairs $n(\omega_{s},\omega_{i})$. The mean spectral density $n$
corresponding to the quantum state $|\psi\rangle$ is defined by the
formula
\begin{equation}  
 n(\omega_{s},\omega_{i})=\langle\langle\psi|\hat{a}_{s}^{\dagger}(\omega_{s}
  )\hat{a}_{s}(\omega_{s})\hat{a}_{i}^{\dagger}(\omega_{i})\hat{a}_{i}(\omega_{i}
  )|\psi\rangle\rangle_{{\rm av}},
\label{11}
\end{equation}
where the symbol $\langle\rangle_{{\rm av}}$ means stochastic
averaging over an ensemble of geometric configurations of an RPS.
Assuming the quantum state $|\psi\rangle$ written in Eq.~(\ref{4})
we arrive at the formula:
\begin{eqnarray}  
 n(\omega_{s},\omega_{i}) & = & |g(\omega_{s},\omega_{i})|^{2}|E_{p}^{(+)}(\omega_{s}+\omega_{i})|^{2}\nonumber \\
 &  & \mbox{}\times\langle|F(\Delta k(\omega_{s},\omega_{i}))|^{2}\rangle_{{\rm
  av}}.
\label{12}
\end{eqnarray}
Spectrum $S_{s}$ of, e.g., the signal field and photon-pair generation
rate $N$ can then be derived using the expressions:
\begin{eqnarray}  
 S_{s} & = & \hbar\omega_{s}\int d\omega_{i}n(\omega_{s},\omega_{i}),\label{13}\\
 N & = & \int d\omega_{s}\int
 d\omega_{i}n(\omega_{s},\omega_{i}).
\label{14}
\end{eqnarray}
The averaged squared modulus of the phase-matching function $F$ as
determined by the formula in Eq.~(\ref{10}) can be written in the
form:
\begin{eqnarray}  
 \langle|F(\Delta k)|^{2}\rangle_{{\rm av}} & = & \frac{4}{\Delta
  k^{2}}\left((N_{L}+1)\frac{1-|H(\delta k)|^{2}}{|1-H(\delta
  k)|^{2}}\right.\nonumber\\
 &  & \hspace{-2cm}\left.\mbox{}-\left[\frac{H(\delta k)[1-H(\delta
  k)^{N_{L}+1}]}{[1-H(\delta k)]^{2}}+{\rm c.c.}\right]\right),
\label{15}
\end{eqnarray}
 $\delta k(\omega_{s},\omega_{i})=\Delta k(\omega_{s},\omega_{i})-\Delta k_{0}$.
Symbol ${\rm c.c.}$ replaces the complex-conjugated term. Function
$H(\delta k)$ occurring in Eq.~(\ref{15}) is defined as:
\begin{eqnarray}   
 H(\delta k) & = & \exp[i\delta kl_{0}]G(\Delta k_{0}+\delta k),\nonumber \\
 G(\Delta k) & = & \exp\left(-\frac{\sigma^{2}\Delta k^{2}}{4}\right).
\label{16}
\end{eqnarray}

The averaged squared modulus $\langle|F(\Delta
k)|^{2}\rangle_{{\rm av}}$ of phase-matching function given in
Eq.~(\ref{15}) determines the averaged spectral density $n$ and
behaves as follows. It holds that $|H|\leq 1$ and $|H|=1$ for a
fully ordered structure. If $\delta k=0$ in a fully ordered
structure, $H$ is real ($H=1$) and the averaged squared
phase-matching function $\langle|F(\Delta k)|^{2}\rangle_{{\rm
av}}$ reaches its maximum value $4(N_{L}+1)^{2}$. Nonzero phase
mismatch $\delta k$ shifts $H$ into the complex plane which
results in lower values of the mean value $\langle|F(\Delta
k)|^{2}\rangle_{{\rm av}}$. The larger the $\delta k$, the smaller
the mean value $\langle|F(\Delta k)|^{2}\rangle_{{\rm av}}$.
Inspection of the formula for $H$ in Eq.~(\ref{16}) also shows
that the larger the value of the basic layer length $l_{0}$ the
faster the decrease of mean values $\langle|F(\Delta
k)|^{2}\rangle_{{\rm av}}$ for given $\delta k$. According to the
formula in Eq.~(\ref{16}) the larger the standard deviation
$\sigma$ of a random structure the smaller the value of $|H|$. The
decrease of values of $|H|$ results in an increase of the range of
values of the phase mismatch $\delta k$ in which the averaged
squared modulus $\langle|F(\Delta k)|^{2}\rangle_{{\rm av}}$ of
phase-matching function attains non-negligible values {[}see the
formula in Eq.~(\ref{15}){]}.

The formula for averaged squared modulus $\langle|F(\Delta
k)|^{2}\rangle_{{\rm av}}$ of phase-matching function in
Eq.~(\ref{15}) can be substantially simplified under the
assumption $\sigma^{2}(\Delta k_{0})^{2}N_{L}/2\gg1$:
\begin{eqnarray}  
 \langle|F(\Delta k)|^{2}\rangle_{{\rm av}} & = & \frac{2N_{L}}{(\Delta k_{0}+\delta k)^{2}}\nonumber \\
 &  & \hspace{-3cm}\mbox{}\times\frac{1-G(\Delta k_{0}+\delta k)}{1-2G(\Delta
 k_{0}+\delta k)\cos(\delta kl_{0})+G(\Delta k_{0}+\delta k)^{2}}.
\label{17}
\end{eqnarray}
Increasing values of phase mismatch $\delta k$ lead to greater
values of the denominator in the fraction on the r.h.s. of
Eq.~(\ref{17}) that result in the decrease of values of the
averaged squared modulus $\langle|F(\Delta k)|^{2}\rangle_{{\rm
av}}$ of phase-matching function. On the other hand, increasing
values of deviation $\sigma$ weaken this behavior.

For comparison, we consider another type of RPS defined such that
$z_{n}=-L+nl_{0}+\delta l_{n}$ where $\delta l_{n}$ is a random
declination of the $n$-th boundary. These `weakly-random'
structures are more ordered compared to those considered earlier
because the change of length of an $n$-th domain is compensated by
the change in length of an $ (n+1) $-th domain. The averaged
squared modulus $\langle|F^{{\rm w-r}}(\Delta k)|^{2}\rangle_{{\rm
av}}$ of phase-matching function can be derived in this case as
follows:
\begin{eqnarray}  
 \langle|F^{{\rm w-r}}(\Delta k)|^{2}\rangle_{{\rm av}} & = & \frac{4}{(\Delta
  k)^{2}}\Biggl\{ N_{L}+1 \nonumber \\
 &  & \hspace{-3cm}\mbox{} + |G(\Delta k_{0}+\delta
 k)|^{2}\left[\frac{\exp(i\delta kl_{0})}{1-\exp(i\delta kl_{0})}\right.\nonumber\\
 &  & \hspace{-3cm}\mbox{}\times\left.\left(N_{L}-\exp(i\delta
  kl_{0})\frac{1-\exp[i\delta kl_{0}N_{L}]}{1-\exp(i\delta
  kl_{0})}\right)+{\rm c.c.}\right]\Biggr\}. \nonumber \\
 & &
\label{18}
\end{eqnarray}
Disorder of the boundary positions manifests itself as a filter
for the averaged squared modulus $\langle|F^{\rm w-r}(\Delta
k)|^{2}\rangle_{{\rm av}}$ of phase-matching function, as evident
from the expression in Eq.~(\ref{18}). This leads to spectral
filtering of the spectral density $n$. This behavior is
qualitatively different from that observed in RPS as described by
the formula in Eq.~(\ref{15}) indicating broadening of the
spectral density $ n $ with increasing values of the deviation
$\sigma$.

Spectral broadening is the most interesting feature of ordered
CPPS that we consider here for comparison. Positions of boundaries
in these structures are described by the formula
$z_{n}=-L+nl_{0}+\zeta'(n-N_{L}/2)^{2}l_{0}^{2}$,
$\zeta'=\zeta/\Delta k_{0}$, and $\zeta$ denotes chirping
parameter. The phase-matching function $F^{{\rm chirp}}(\Delta k)$
then takes the form \cite{Harris2007}:
\begin{eqnarray}  
 &  & F^{{\rm chirp}}(\Delta k)=\frac{2\sqrt{\pi}}{\sqrt{i\Delta k^{3}\zeta'}\,
  l_{0}}\exp(i\Delta kN_{L}l_{0}/2) \nonumber \\
 &  & \times\exp\left(-\frac{i\delta k^{2}}{4\Delta k\zeta'}\right)\left[{\rm
  erf}(f(N_{L}/2))-{\rm erf}(f(-N_{L}/2))\right],\nonumber \\
 & &
 \label{19} \\
 &  & f(x)=\frac{\sqrt{-i}}{2}\left[\sqrt{\zeta'(\Delta k_{0}+\delta
  k)}xl_{0}+\frac{\delta k}{\sqrt{\zeta'(\Delta k_{0}+\delta
  k)}}\right].\nonumber
\end{eqnarray}
The error function ${\rm erf}$ is defined as ${\rm
erf}(x)=2/\sqrt{\pi}\int_{0}^{x}\exp(-y^{2})dy$. Detailed
inspection of the formula in Eq.~(\ref{19}) reveals that the
larger the value of chirping parameter $\zeta'$ the broader the
phase-matching function $F^{{\rm chirp}}(\Delta k)$.

As an example, we consider spectrally degenerate (nearly)
collinear down-conversion from periodically-poled LiNbO$_{3}$
pumped at the wavelength $\lambda_{p}^{0}=775$~nm by a cw laser
beam. The signal and idler photons thus occur at the fiber-optics
communication wavelength $\lambda_{s}=\lambda_{i}=1.55~\mu$m. The
crystal optical axis is perpendicular to the fields' propagation
direction and is parallel to the vertical direction. All fields
are vertically polarized and so the largest element
$\chi_{33}^{(2)}$ of the susceptibility tensor is used. The
natural phase mismatch for this configuration is compensated by
the basic domain length $l_{0}$ equal to 9.51535~$\mu$m. A
structure composed of $N_{L}=700$ layers is roughly 6.5~mm long
and typically delivers $4\times10^{6}$ photon pairs per 100~mW of
pumping in case of weakly random positions of boundaries (small
values of variance $\sigma$).

\begin{figure}  
 (a) \includegraphics[scale=0.3]{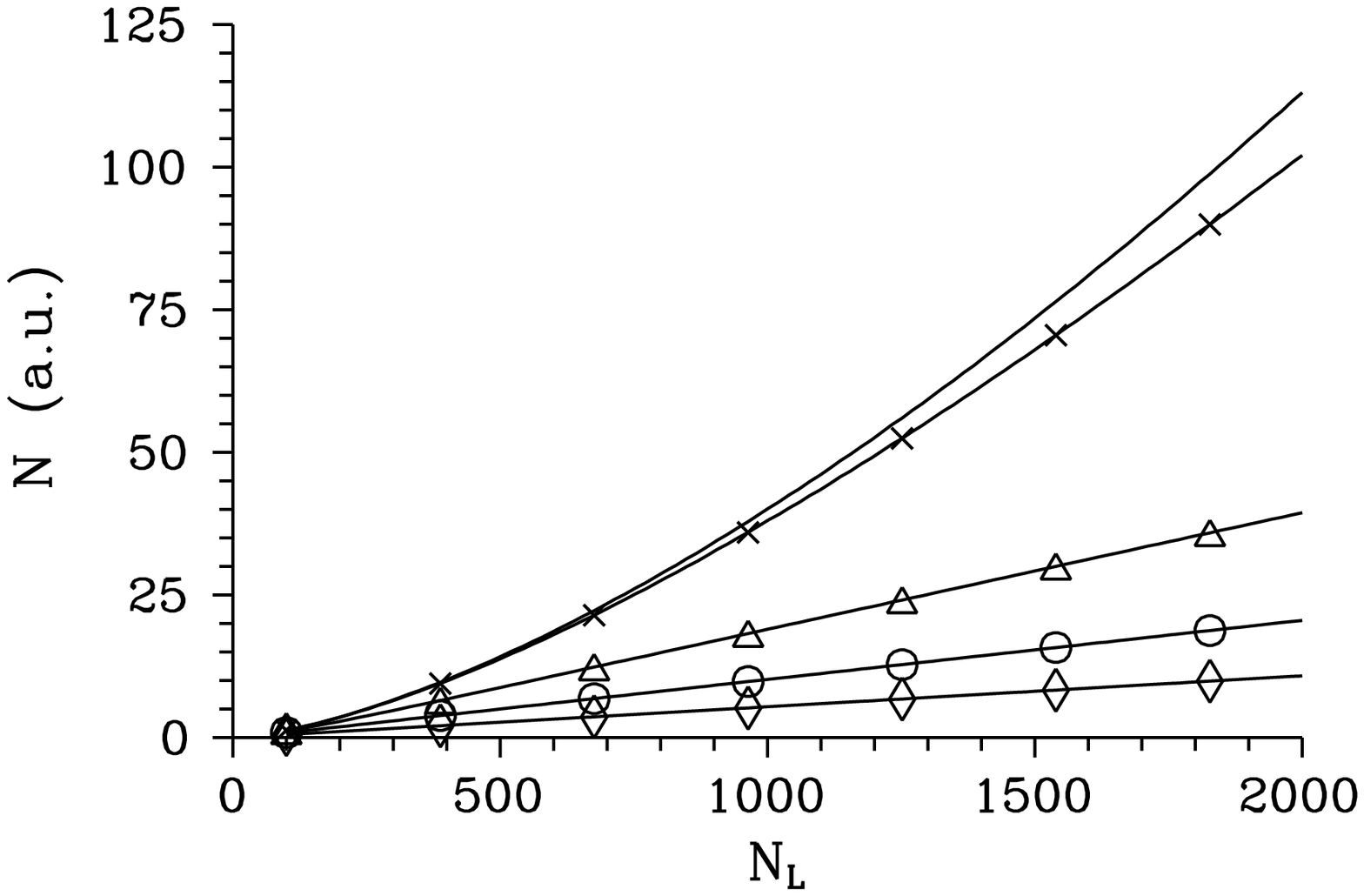}

 \vspace{7mm}
 (b) \includegraphics[scale=0.3]{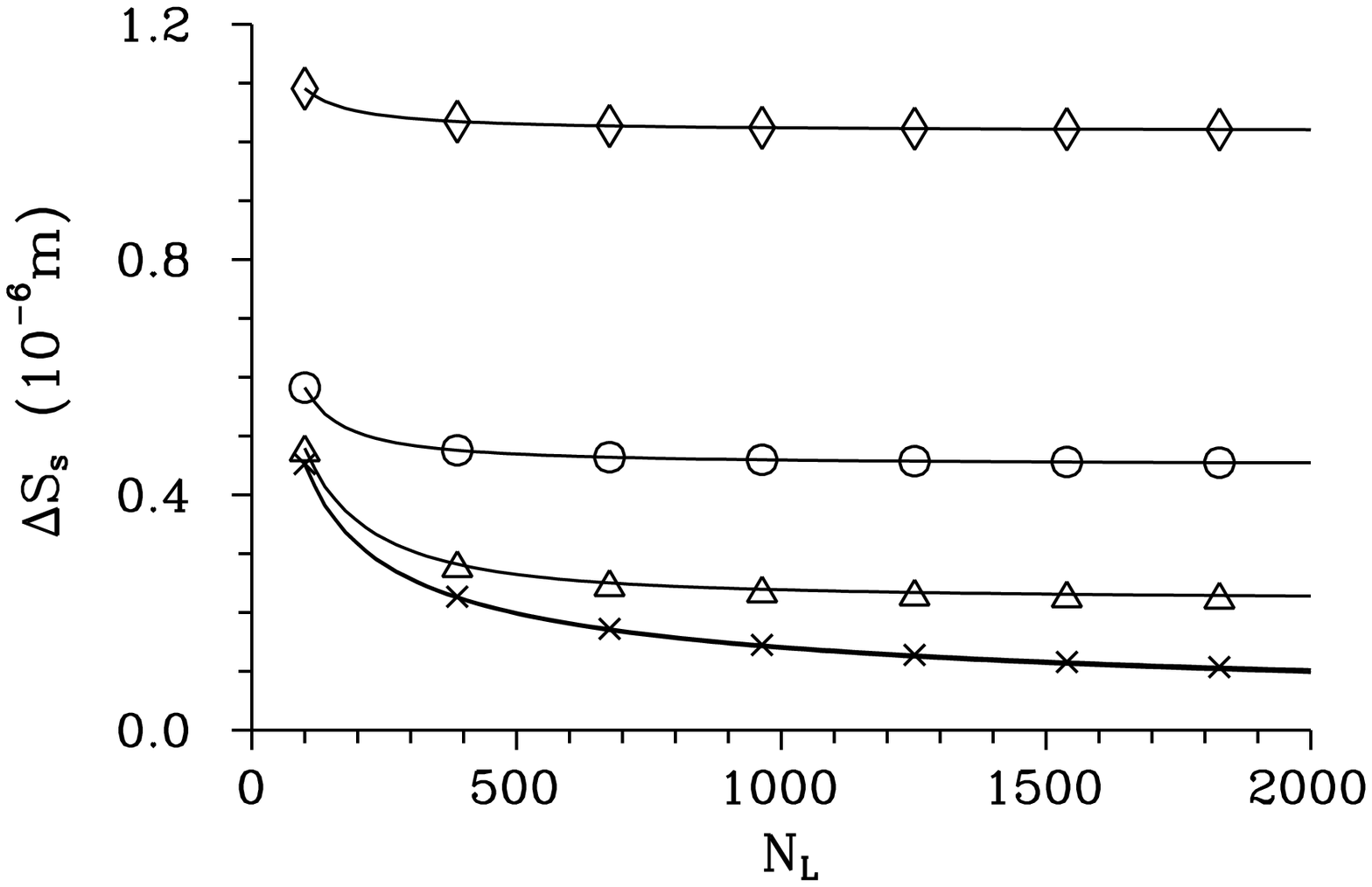}

 \caption{(a) Photon-pair generation rate $N$ and (b) signal-field
  spectral width $ \Delta S_{s}$ (FWHM) as functions of the number $N_{L}$
  of domains for an ensemble of RPSs with standard
  deviation $\sigma$ equal to 0~m (solid curve), 0.1~$\times
  10^{-6}$m (solid curve with $\times$), 0.5~$\times 10^{-6}$m
  (solid curve with $\triangle$), 1~$\times 10^{-6}$m (solid curve
  with $\circ$), and 2~$\times 10^{-6}$m (solid curve with
  $\diamond$); $ {\rm a.u.} $ stands for arbitrary units.}
\label{fig1}
\end{figure}

The most striking feature of RPSs is that the photon-pair
generation rate $N$ increases linearly with the number $N_{L}$ of
domains, independently of the standard deviation $\sigma$ of the
random positions of boundaries [see Fig.~\ref{fig1}(a)]. Standard
deviation $\sigma$ plays the central role in the determination of
spectral widths $ \Delta S_{s} $ and $ \Delta S_{i} $ of the
signal and idler fields. The larger the value of deviation
$\sigma$ the broader the signal- and idler-field spectra $S_{s}$
and $S_{i}$, as documented in Fig.~\ref{fig1}b. This behavior is
easily understandable because structures generated with larger
values of $\sigma$ have statistically a broader spatial spectrum
of the $\chi^{(2)}(z)$ modulation which gives more freedom for the
fulfillment of quasi-phase-matching conditions. It holds that the
broader the signal- and idler-field spectra $S_{s}$ and $S_{i}$
the smaller the photon-pair generation rate $N$ [compare
Figs.~\ref{fig1}(a) and (b)]. It reflects the fact that
constructive interference of fields from different domains is
enhanced in the area outside the central frequencies
$\omega_{s}^{0}$ and $\omega_{i}^{0}$ whereas this interference is
weaken in the area around the central frequencies.

\begin{figure}  
 \includegraphics[scale=0.3]{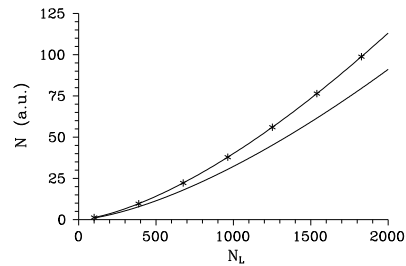}

 \caption{Photon-pair generation rate $N$ as it depends on the
 number $N_{L}$ of domains for an ensemble of `weakly-random'
 structures with standard deviation $\sigma$ equal to 0~m (solid
 curve with $\ast$) and 2~$\times 10^{-6}$m (solid curve).}
\label{fig2}
\end{figure}

The photon-pair generation rate $N$ increases roughly linearly
with the number $N_{L}$ of domains also in the case of
`weakly-random' structures described by the averaged squared
modulus $\langle|F^{{\rm w-r}}(\Delta k)|^{2}\rangle_{{\rm av}}$
of phase-matching function in Eq.~(\ref{18}), as shown in
Fig.~\ref{fig2}. The greater the standard deviation $\sigma$ the
smaller the photon-pair generation rate $N$. As for the
signal-field spectral width $ \Delta S_{s} $ its values do not
practically depend on the variance $\sigma$.

\begin{figure} 
 (a) \includegraphics[scale=0.3]{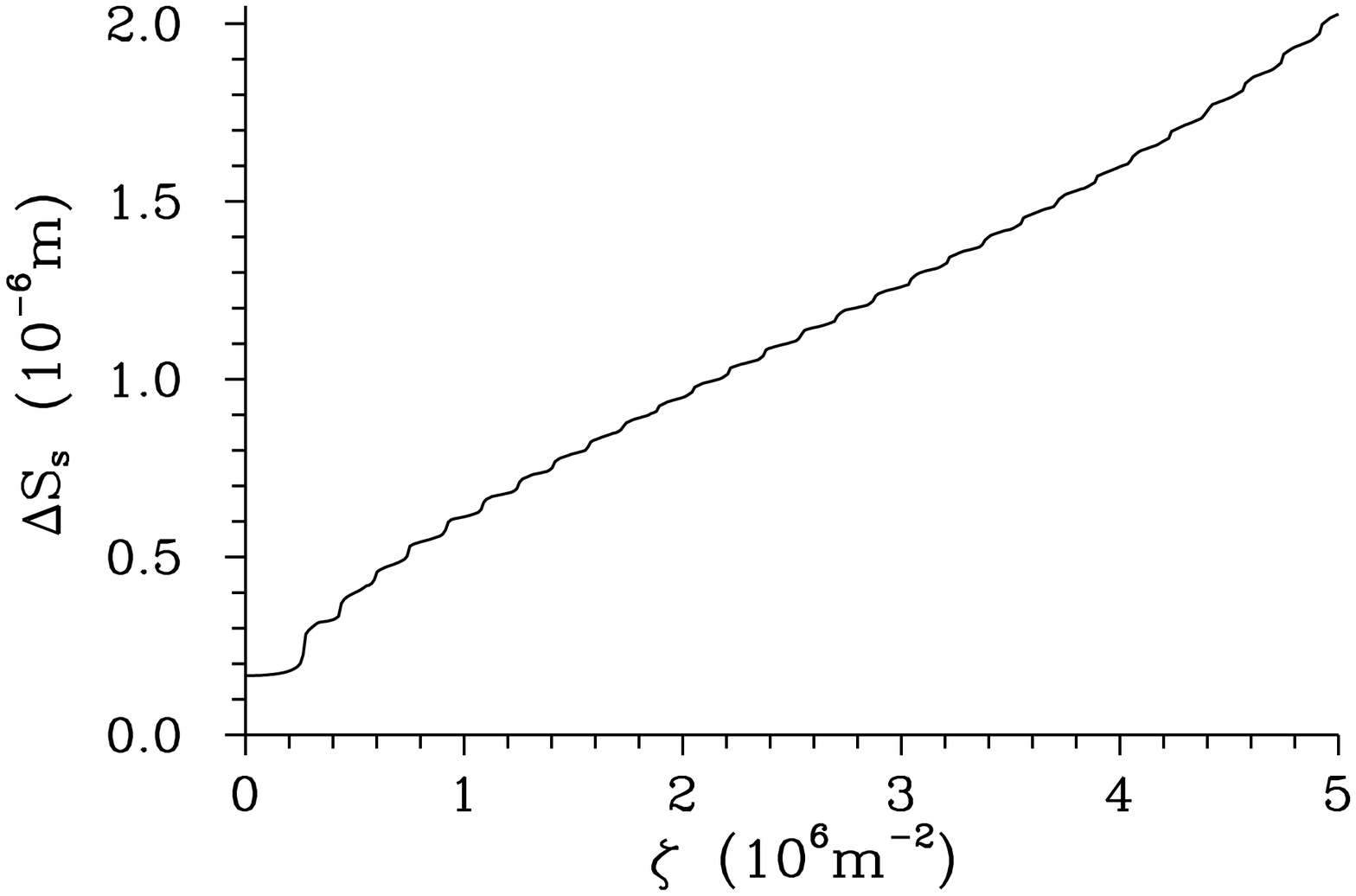}

\vspace{7mm}
 (b) \includegraphics[scale=0.3]{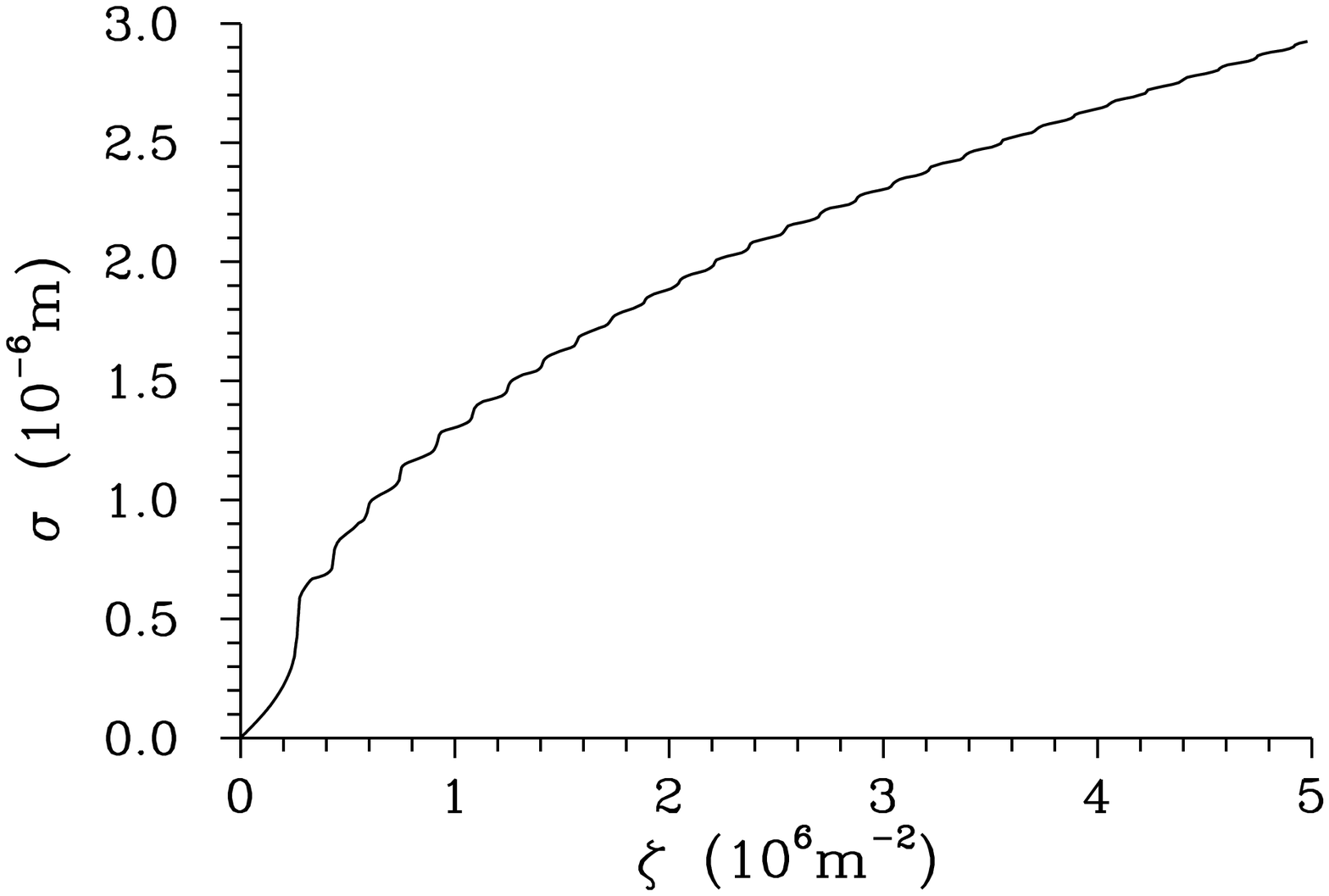}

\vspace{7mm}
 (c) \includegraphics[scale=0.3]{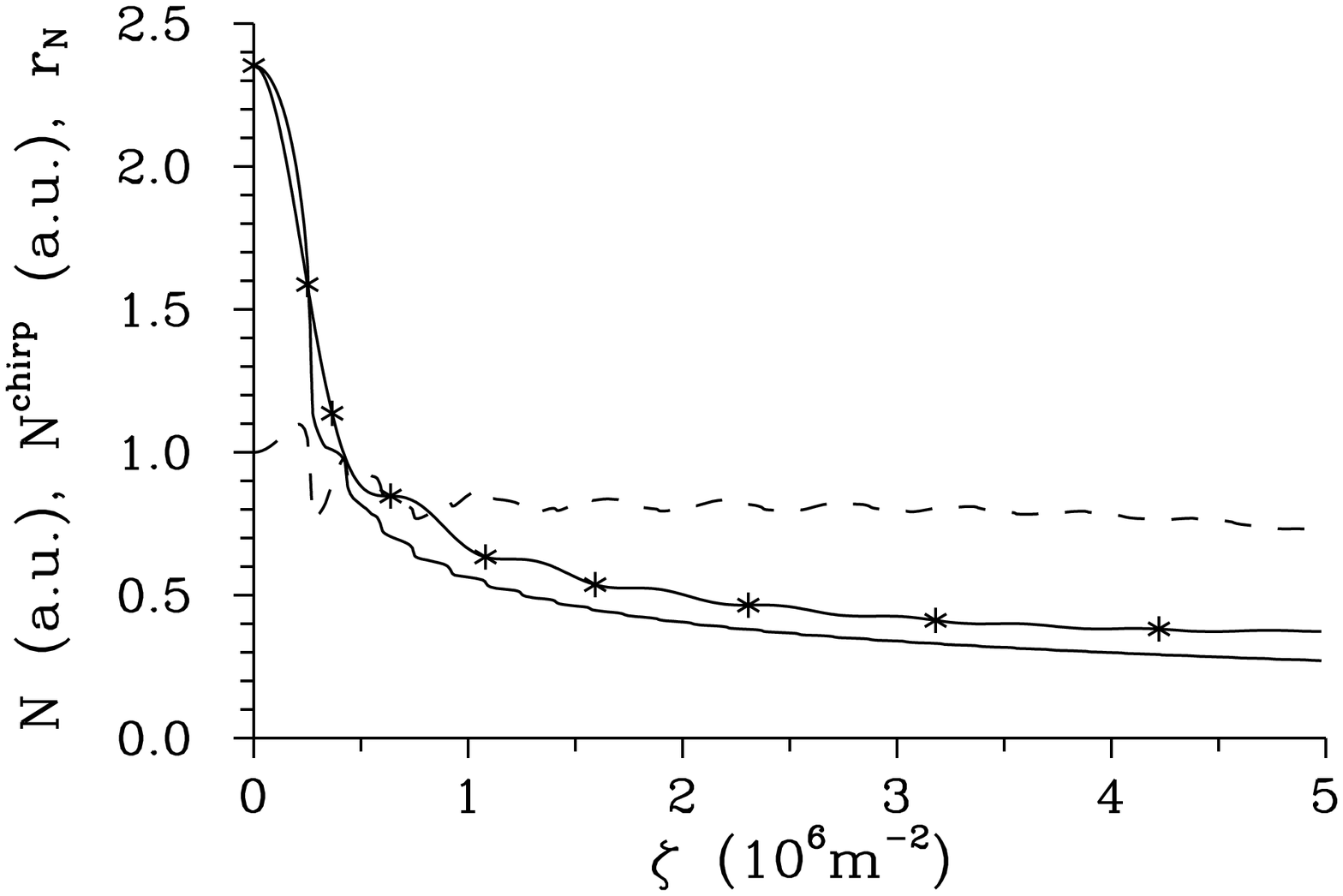}

 \caption{(a) Signal-field spectral width $ \Delta S_{s}$ (FWHM) as a
  function of chirping parameter $\zeta$. (b) Transformation curve
  between the standard deviation $\sigma$ and chirping parameter
  $\zeta$ assuming the same spectral widths $ \Delta S_{s}$. (c) Photon-pair
  generation rate for chirped ($N^{{\rm chirp}}$, solid curve with
  $\ast$) and random ($N$, solid curve) structures and their ratio
  $r_{N}$ ($r_{N}=N/N^{{\rm chirp}}$, dashed curve) as functions of
  chirping parameter $\zeta$; $N_{L}=700$.}
\label{fig3}
\end{figure}

The behavior of photon-pair generation as observed in RPSs can
also be found in ordered CPPSs. Also here photon-pair generation
rate $N$ is linearly proportional to the number $ N_L $ of domains
and spectral widths $ \Delta S_{s}$ and $ \Delta S_{i}$ increase
with increasing chirping parameter $\zeta$. The main result of our
analysis is that this similarity is both qualitative and
quantitative. For any value of the chirping parameter $\zeta$
there exists a value of the standard deviation $\sigma$ such that
spectral widths $ \Delta S_{s} $ and $ S\Delta _{i} $ of the
generated signal and idler fields are the same. Moreover (and a
bit surprisingly), also photon-pair generation rates $N$ are
comparable. This is illustrated in Fig.~\ref{fig3} for a chirped
structure with $N_{L}=700$ domains. Its signal-field spectrum
$S_{s}$ is extraordinarily broad (larger that 1~$\mu$m) for
sufficiently large values of the chirping parameter $\zeta$ [see
Fig.~\ref{fig3}(a)]. Signal-field spectra $S_{s}$ of the same
width can also be generated from RPSs with sufficiently large
randomness (i.e., having large values of the deviation $\sigma$).
Values of the standard deviation $\sigma$ corresponding to the
values of chirping parameter $\zeta$ are plotted in
Fig.~\ref{fig3}(b). Photon-pair generation rates $N$ for RPSs and
CPPSs are drawn for comparison in Fig.~\ref{fig3}(c) in this case.
Whereas CPPSs give better photon-pair generation rates $N$ for
larger values of chirping parameter $\zeta$, RPSs even slightly
overcome on average CPPSs for smaller values of $\zeta$. Moreover,
the signal-field spectra $S_{s}$ of individual realizations may
even be broader which results in sharper temporal features. On the
other hand, these spectra are typically composed of many local
peaks (see Fig.~4). RPSs thus represent an alternative broadband
and efficient source of photon pairs with properties comparable to
CPPSs. We note, that histograms of domain lengths corresponding to
RPSs are broader compared to those characterizing CPPSs.

\begin{figure} 
 \includegraphics[scale=0.21]{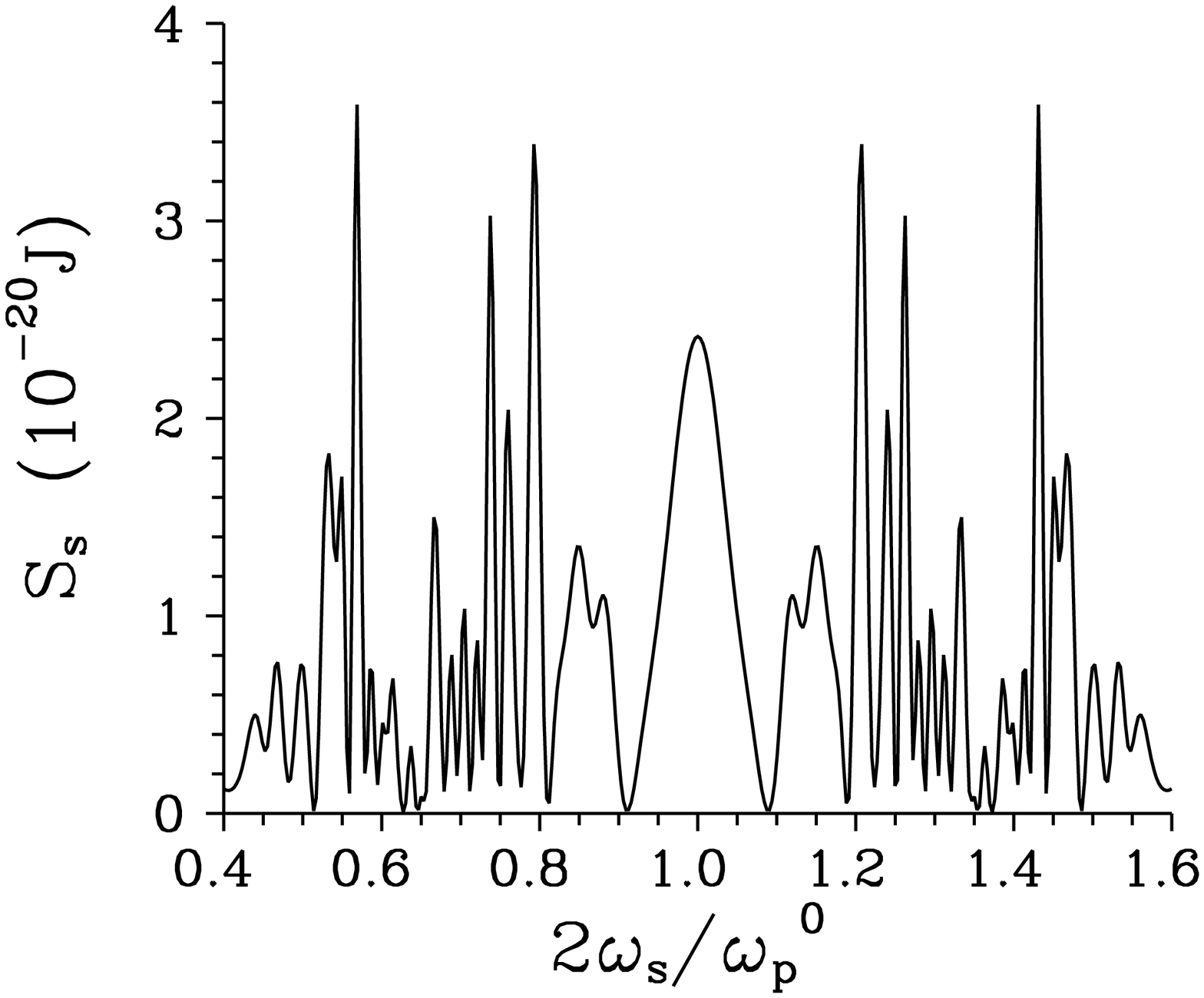} \hspace{2mm} \includegraphics[scale=0.21]{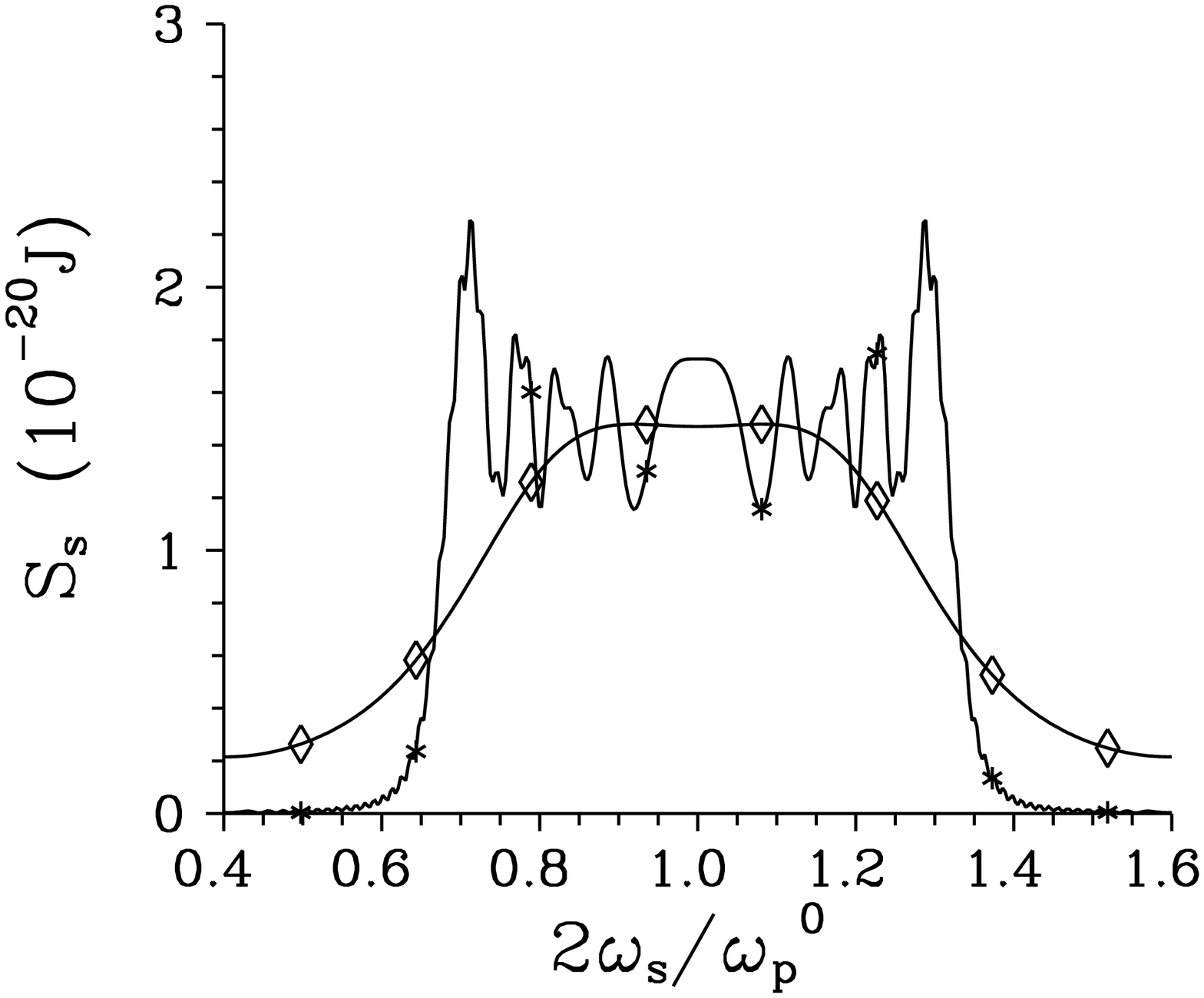}

 (a) \hspace{3.5cm} (b)

 \caption{Signal-field spectrum $S_{s}$ for (a) one typical realization of
  RPS (solid curve) and (b) CPPS (solid curve with $\ast$)
  and an ensemble of RPSs (solid
  curve with $\diamond$). Spectra $S_{s}$ are normalized such that
  one photon is emitted; $\sigma=2.1$~$\times 10^{-6}$m, $\zeta=2.5\times 10^{6}$m$ {}^{-2}$,
  $N_{L}=700$.}
\label{fig4}
\end{figure}

\begin{figure}  
 (a) \includegraphics[scale=0.3]{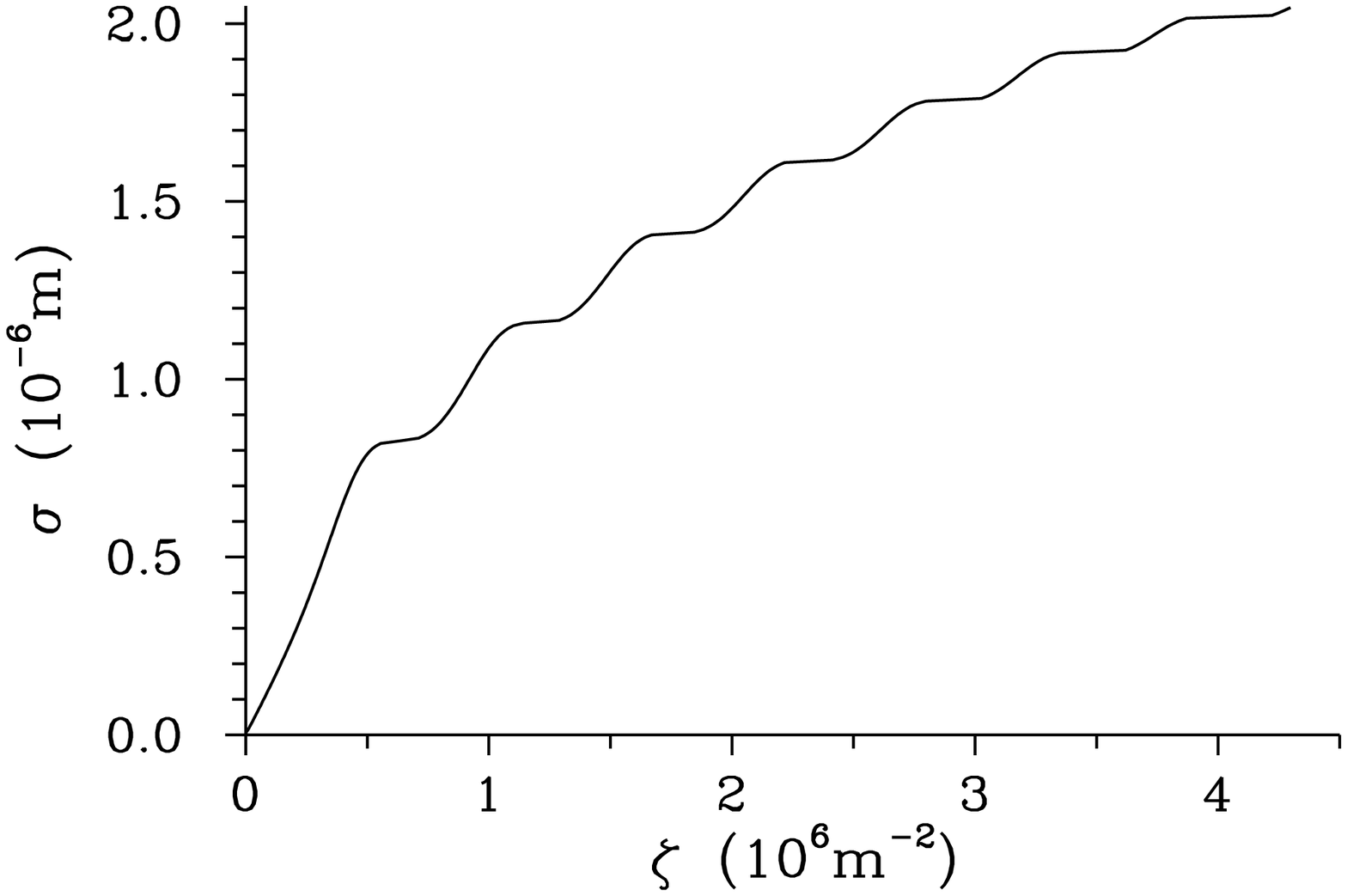}

 \vspace{7mm}
 (b) \includegraphics[scale=0.3]{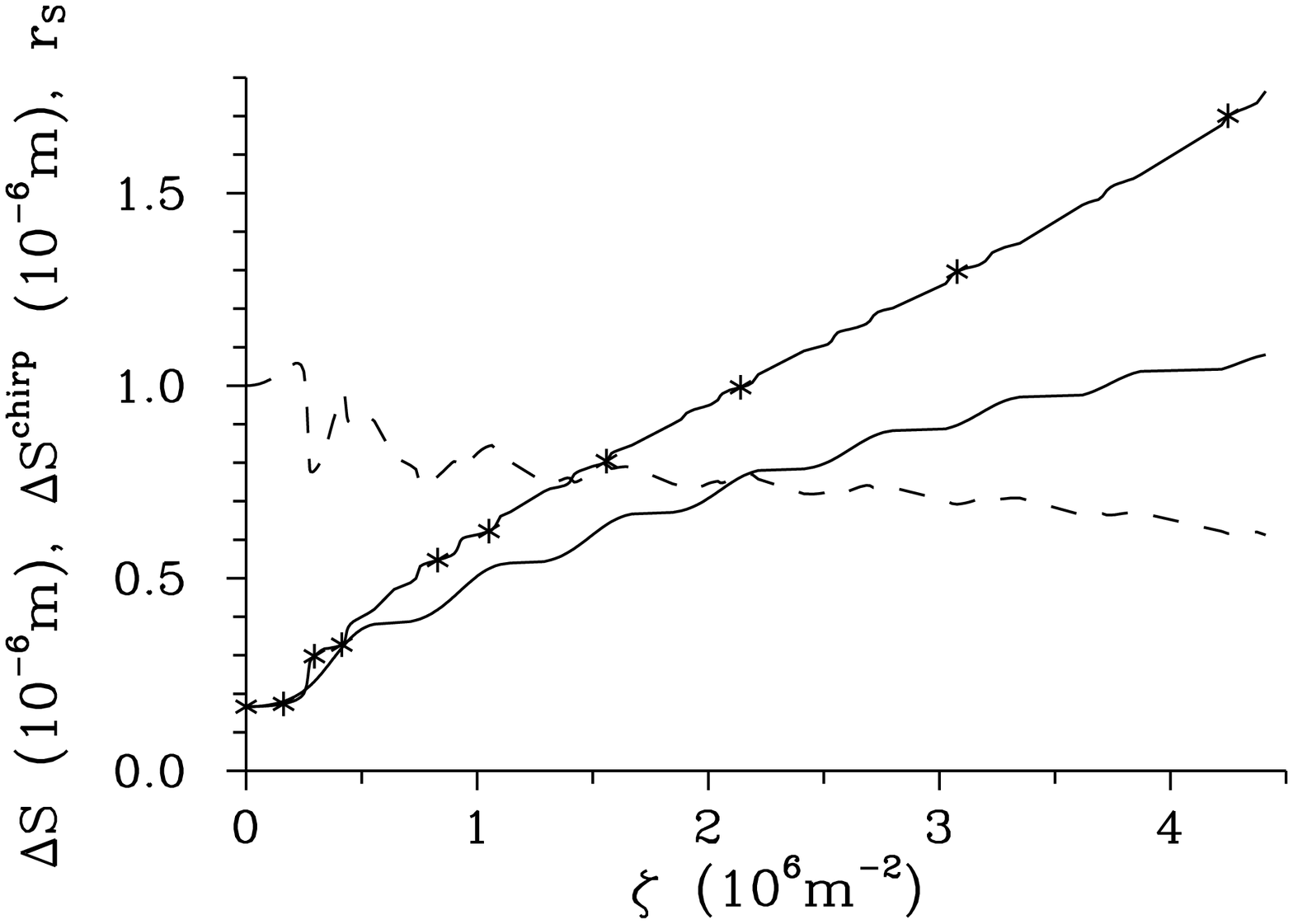}

 \caption{(a) Transformation curve between the standard deviation $\sigma$ of
  RPSs and chirping parameter $\zeta$ assuming the same
  photon-pair generation rates $N$. (b) Signal-field spectral widths
  (FWHM) for random ($ \Delta S_{s} $, solid curve) and chirped ($ \Delta S_{s}^{{\rm chirp}} $,
  solid curve with $\ast$) structures and their ratio $r_{S}$ ($r_{S}=S_{s}/S_{s}^{{\rm chirp}}$,
  dashed curve) as functions of chirping parameter $\zeta$; $N_{L}=700$.}
\label{fig5}
\end{figure}

Alternatively, RPSs and CPPSs can be compared under the
requirement of equal photon-pair generation rates $N$. Values of
the photon-pair generation rate $N$ decrease with the increasing
values of chirping parameter $\zeta$ [see Fig.~\ref{fig3}(c)].
Transformation curve between standard deviation $\sigma$ and
chirping parameter $\zeta$ stemming from the requirement of equal
generation rates $N$ is monotonous and is plotted in
Fig.~\ref{fig5}(a) in the considered case. The corresponding
signal-field widths $ \Delta S_{s}$ plotted in Fig.~\ref{fig5}(b)
reveal that CPPSs provide broader spectra for the most of values
of chirping parameter $\zeta$. However, the difference in spectral
widths in CPPSs and RPSs is not dramatic.

\begin{figure} 
 (a)\includegraphics[scale=0.85]{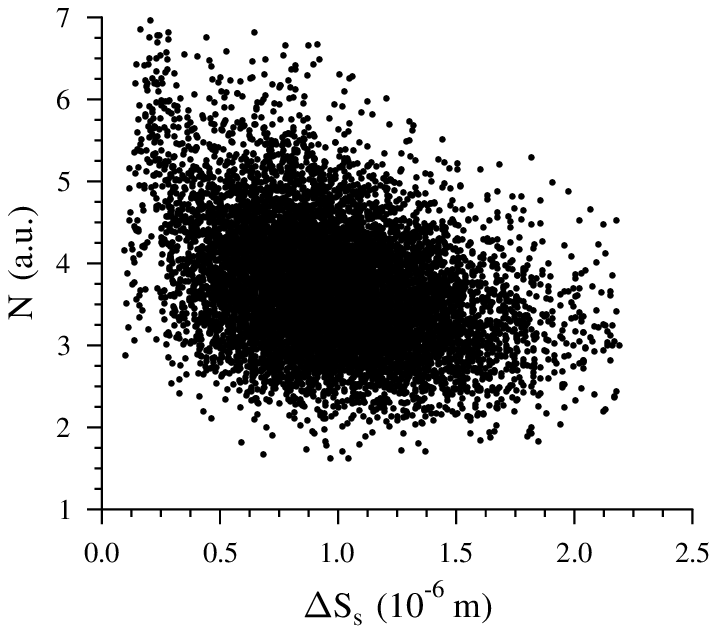}

 (b) \includegraphics[scale=0.7]{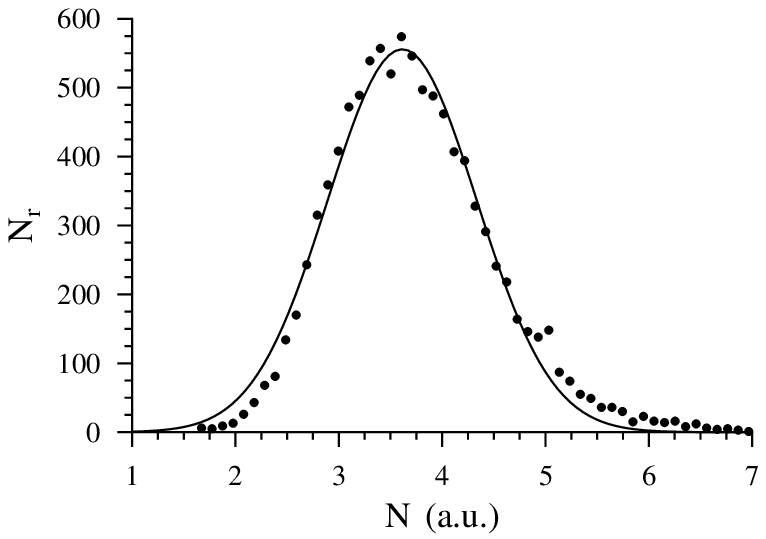}

 \vspace{7mm}
 (c) \includegraphics[scale=0.7]{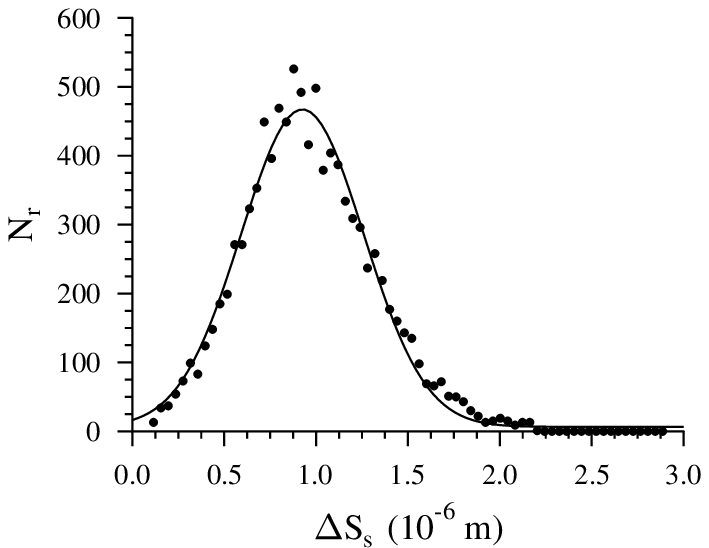}

 \caption{(a) Photon-pair generation rates $N$ and signal-field
  spectral widths $\Delta S_{s}$ for 10000 realizations of RPSs
  (each realization is depicted as a point). (b), (c)
  Histograms of rates $ N $ (b) and widths $\Delta S_{s}$ (c); $
  N_r $ gives the number of samples with given properties. Solid
  curves in (b) and (c) are best-fit Gaussian curves;
  $\sigma=2.1$ $\times 10^{-6} $m, $N_{L}=700$.}
\label{fig6}
\end{figure}

The above presented results for RPSs represent an ensemble
average. In practical applications, properties of individual
realizations of a given RPS are naturally important. In
Fig.~\ref{fig6}, we show generation rates $N$ and signal-field
widths $ \Delta S_{s}$ for 10000 realizations of the RPS.
Histograms of the generation rates $N$ and signal-field widths $
\Delta S_{s} $ plotted in Figs.~\ref{fig6}(b) and \ref{fig6}(c)
are close to Gaussian distributions, in accordance with the
central limiting theorem.

\begin{figure}    
 \includegraphics[scale=0.85]{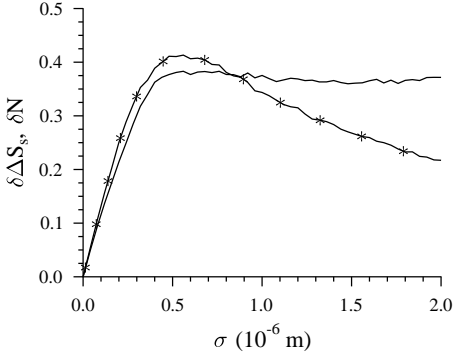}
 \caption{Relative quadratic fluctuations $\delta\Delta S_{s}$ of the signal-field spectral width
 (solid curve) and relative quadratic fluctuations $\delta N$ of the photon-pair generation rate
 (solid curve with $ \ast $) as they depend on standard deviation
  $\zeta$; $N_{L}=700$.}
\label{fig7}
\end{figure}

It holds that the larger the deviation $ \sigma $ the smaller the
relative quadratic fluctuations $\delta N$ and $\delta\Delta
S_{s}$ of photon-pair generation rates $ N $ and signal-field
spectral widths $ \Delta S_{s} $, respectively [$\delta
x=\sqrt{\langle(\Delta x)^{2}\rangle_{{\rm av}}}/\langle
x\rangle_{{\rm av}}$, $\Delta x=x-\langle x\rangle_{{\rm av}} $]
(see Fig.~\ref{fig7}). However, we should note that these relative
fluctuations are quite large and can even approach 40~\%.

\section{Temporal correlations, entanglement time}

There occurs a strong correlation between possible detection times
of a signal photon and its twin from one photon pair because both
photons are generated inside the nonlinear medium at one instant.
A finite distance between the detection times of both photons is a
consequence of dispersion properties of the nonlinear medium
through which both photons at different frequencies propagate
before they leave the crystal. Temporal correlations of the signal
and idler photons can be conveniently described using a two-photon
temporal amplitude ${\cal A}$ defined as:
\begin{eqnarray}  
 {\cal A}(t_{s},t_{i})=\langle{\rm
 vac}|\hat{E}_{s}^{(+)}(t_{s})\hat{E}_{i}^{(+)}(t_{i})|\psi\rangle.
\label{20}
\end{eqnarray}
This amplitude ${\cal A}(t_{s},t_{i})$ gives the probability
amplitude of detecting a signal photon at time $t_{s}$ and an
idler photon at time $t_{i}$.

The simplest experimental method for the determination of a
typical constant characterizing temporal width of the two-photon
detection window (entanglement time) uses a Hong-Ou-Mandel
interferometer. In this interferometer, the signal and idler
fields mutually interfere on a beam-splitter and photons leaving
the beam-splitter at different output ports are subsequently
detected in a coincidence-count measurement. The coincidence-count
rate $R_{n}$ depends on a mutual time delay $\tau$ introduced
between the signal and idler photons. It can be shown that
temporal extension of the interference part in the
coincidence-count rate $R_{n}$ is proportional to entanglement
time under certain conditions. The coincidence-count rate $R_{n}$
as a function of relative time delay $\tau$ is described by the
following formula:
\begin{equation}  
 R_{n}(\tau)=1-\varrho(\tau),
\label{21}
\end{equation}
 where
\begin{eqnarray}  
 \rho(\tau) & = & \frac{1}{2R_{0}}\int_{-\infty}^{\infty}dt_{1}\int_{-\infty}^{\infty}dt_{2}\nonumber \\
 &  & {\rm Re}\left[\langle{\cal A}(t_{1},t_{2}-\tau){\cal A}^{*}(t_{2},t_{1}-\tau)\rangle_{{\rm av}}\right],
\label{22}\\
 R_{0} & = & \frac{1}{2}\int_{-\infty}^{\infty}dt_{1}\int_{-\infty}^{\infty}dt_{2}\langle|{
  \cal A}(t_{1},t_{2})|^{2}\rangle_{{\rm av}}.
\label{23}
\end{eqnarray}
Inserting Eqs.~(\ref{20}) and (\ref{4}) for the two-photon
temporal amplitude ${\cal A}$ and quantum state $|\psi\rangle$,
respectively, into Eqs.~(\ref{22}) and (\ref{23}) we arrive at the
expressions:
\begin{eqnarray}  
 \rho(\tau) & = & \frac{\pi\hbar^{2}}{4\epsilon_{0}^{2}c^{2}{\cal
  B}^{2}}\frac{1}{R_{0}}{\rm
  Re}\Biggl[\int_{0}^{\infty}d\omega_{s}\,\int_{0}^{\infty}d\omega_{i}\;\omega_{s}
  \omega_{i}\nonumber \\
 &  & \hspace{-12mm}\langle\Phi(\omega_{s},\omega_{i})
  \Phi^{*}(\omega_{i},\omega_{s})\rangle_{{\rm av}}\exp(i\omega_{i}\tau)\exp(-i\omega_{s}\tau)\Biggr],\nonumber \\
  \label{24}\\
 R_{0} & = & \frac{\pi\hbar^{2}}{4\epsilon_{0}^{2}c^{2}{\cal
  B}^{2}}\int_{0}^{\infty}d\omega_{s}\,\int_{0}^{\infty}d\omega_{i}\,\;\omega_{s}
  \omega_{i}\langle|\Phi(\omega_{s},\omega_{i})|^{2}\rangle_{{\rm av}}.\nonumber\\
\label{25}
\end{eqnarray}

For simplicity, we further assume cw-pumping with amplitude $
\xi_p $ at frequency $\omega_{p}^{0}$, i.e.
$E_{p}^{(+)}(\omega_{p})=\xi_{p}\delta(\omega_{p}-\omega_{p}^{0})$.
Formulas in Eqs.~(\ref{24}) and (\ref{25}) can be simplified in
this case and recast into the following form:
\begin{eqnarray}  
 \rho(\tau) & = & \frac{\hbar^{2}}{8\epsilon_{0}^{2}c^{2}{\cal B}^{2}}
  \frac{|\xi_{p}|^{2}}{R_{0}}{\rm Re}\Biggl[\exp(i\omega_{p}^{0}\tau)
  \int_{0}^{\infty}d\omega_{s}\nonumber \\
 &  & \hspace{0mm}\omega_{s}(\omega_{p}^{0}-\omega_{s})
  |g(\omega_{s},\omega_{p}^{0}-\omega_{s})|^{2}\exp(-2i\omega_{s}\tau)\nonumber \\
 &  & \hspace{0mm}\times{\cal F}\left(\Delta k(\omega_{s},\omega_{p}^{0}-\omega_{s}),
  \Delta k(\omega_{p}^{0}-\omega_{s},\omega_{s})\right)\Biggr],\nonumber \\
\label{26}\\
 R_{0} & = & \frac{\hbar^{2}|\xi_{p}|^{2}}{8\epsilon_{0}^{2}c^{2}{\cal B}^{2}}
  \int_{0}^{\infty}d\omega_{s}\omega_{s}(\omega_{p}^{0}-\omega_{s})
  |g(\omega_{s},\omega_{p}^{0}-\omega_{s})|^{2}\nonumber \\
 &  & \hspace{-5mm}\times\left|{\cal F}\left(\Delta
  k(\omega_{s},\omega_{p}^{0}-\omega_{s}),\Delta
  k(\omega_{s},\omega_{p}^{0}-\omega_{s})\right)\right|^{2};\nonumber
  \nonumber \\
\label{27}
\end{eqnarray}
$g_{0}\equiv g(\omega_{s}^{0},\omega_{i}^{0})$. Function ${\cal
F}$ introduced in Eqs.~(\ref{26}) and (\ref{27}) incorporates
phase-matching conditions into the description of temporal
properties of photon pairs and is defined according to the
formula:
\begin{equation} 
 {\cal F}(\Delta k,\Delta k')=\langle
  F(\Delta k)F^{*}(\Delta k')\rangle_{{\rm
  av}};
\label{28}
\end{equation}
phase-matching function $F$ has been introduced in Eq.~(\ref{10}).

Considering RPSs with fluctuations of boundaries described by a
Gaussian distribution in Eq.~(\ref{8}) function ${\cal F}$ in
Eq.~(\ref{28}) takes the form:
\begin{eqnarray}  
 {\cal F}(\Delta k,\Delta k') & = & \frac{4}{\Delta k\Delta k'}\tilde{{\cal F}}(\Delta k,\Delta k')\nonumber \\
 &  & \mbox{}\times\exp[-i(\Delta k-\Delta k')N_{L}l_{0}],
\label{29}\\
  \tilde{{\cal F}}(\Delta k,\Delta k') & = & \frac{1-H(Dk)^{N_{L}+1}}{1-H(Dk)}\Biggl[\frac{H(\Delta k)}{H(\Delta k)+H(Dk)}\nonumber \\
 &  & \hspace{-2cm}\times\Biggl(-H(\Delta k)\frac{1-[-H(\Delta k)]^{N_{L}}}{1+H(\Delta k)}\nonumber \\
 &  & \mbox{}-H(Dk)\frac{1-[H(Dk)]^{N_{L}}}{1-H(Dk)}\Biggr)\nonumber \\
 &  & \mbox{}+\left(\Delta k\longleftrightarrow-\Delta k'\right)\Biggr];
\label{30}
\end{eqnarray}
$Dk=\Delta k-\Delta k'$. Function $H$ occurring in Eq.~(\ref{30})
has been defined in Eq.~(\ref{16}). Symbol $(\Delta
k\longleftrightarrow-\Delta k')$ in Eq.~(\ref{30}) replaces the
term that is obtained by the indicated exchange applied to the
preceded term inside the brackets.

Considering `weakly-random' structures, the following form of the
function $\tilde{{\cal F}}$ can be derived:
\begin{eqnarray}  
 \tilde{{\cal F}}^{{\rm w-r}}(\Delta k,\Delta k') & = & G(Dk)\frac{1-\exp[iDkl_{0}(N_{L}+1)]}{1-\exp(iDkl_{0})}\nonumber \\
 &  & \hspace{-2cm}\mbox{}+\Biggl[\frac{G(\Delta k)G(\Delta k')}{1-\exp(-i\Delta k'l_{0})}\Biggl(\frac{1
  -\exp(i\Delta kN_{L}l_{0})}{1-\exp(-i\Delta kl_{0})}\nonumber \\
 &  & \mbox{}+\frac{1-\exp(-iDkN_{L}l_{0})}{1-\exp(-iDkl_{0})}\Biggr)\nonumber \\
 &  & \mbox{}+\left(\Delta k\longleftrightarrow-\Delta k'\right)\Biggr].
\label{31}
\end{eqnarray}

On the other hand, function $\tilde{{\cal F}}^{{\rm chirp}}$
attains a simple form in case of CPPSs:
\begin{equation} 
 \tilde{{\cal F}}^{{\rm chirp}}(\Delta k,\Delta k')=F^{{\rm chirp}}(\Delta k)F^{{\rm chirp}*}(\Delta
  k'),
\label{32}
\end{equation}
where the formula for $F^{{\rm chirp}}$ is written in
Eq.~(\ref{19}).

Characteristics of temporal correlations (correlation time)
between the signal and idler fields can also be obtained from the
measurement of sum-frequency intensity in a nonlinear medium
combining the signal and idler photons and having a sufficiently
high value of $\chi^{(2){\rm sum}}$ nonlinearity. This process
allows us to determine the temporal correlation function $ I^{\rm
sum} $ of intensities of the signal and idler fields. Intensity
$I^{{\rm sum}}$ of the sum-frequency field is given along the
formula:
\begin{eqnarray} 
 I^{{\rm sum}}(\tau) & = & \eta^{{\rm
  sum}}\int_{-\infty}^{\infty}dt\left|\langle{\rm
  vac}|\hat{E}_{s}^{(+)}(t)\hat{E}_{i}^{(+)}(t-\tau)|\psi\rangle\right|^{2},
  \nonumber\\
\label{33}
\end{eqnarray}
where constant $\eta^{{\rm sum}}$ incorporates the value of
$\chi^{(2){\rm sum}}$ nonlinearity and quantum detection
efficiency.

The general formula in Eq.~(\ref{33}) can be recast into the
following form using the expression for function ${\cal F}$ in
Eq.~(\ref{28}):
\begin{eqnarray} 
 I^{{\rm sum}}(\tau) & = & \frac{\eta^{{\rm sum}}\hbar^{2}}{4\varepsilon_{0}^{2}c^{2}{\cal B}^{2}}
  \int_{0}^{\infty}d\omega_{p}|E_{p}^{(+)}(\omega_{p})|^{2}\nonumber \\
 &  & \hspace{-1.3cm}\mbox{}\int_{0}^{\infty}d\omega_{s}\sqrt{\omega_{s}(\omega_{p}-\omega_{s})}
  \int_{0}^{\infty}d\omega'_{s}\sqrt{\omega'_{s}(\omega_{p}-\omega'_{s})}\nonumber \\
 &  & \hspace{-1.3cm}\mbox{}\times g(\omega_{s},\omega_{p}-\omega_{s})
  g^{*}(\omega'_{s},\omega_{p}-\omega'_{s})\exp[-i(\omega_{s}-\omega'_{s})\tau]\nonumber \\
 &  & \hspace{-1.3cm}\mbox{}\times{\cal F}\left(\Delta
  k(\omega_{s},\omega_{p}-\omega_{s}),\Delta k(\omega'_{s},\omega_{p}-\omega'_{s})\right).
\label{34}
\end{eqnarray}
When deriving Eqs.~(\ref{33}) and (\ref{34}) we have assumed that
the nonlinear medium in which sum-frequency generation occurs is
ideally phase matched for frequencies present in the signal and
idler fields.

\begin{figure}  
 \includegraphics[scale=0.3]{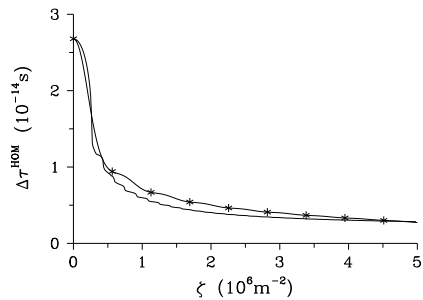}
 \caption{Entanglement time $\Delta\tau^{{\rm HOM}}$ (FWHM) as a function of
  chirping parameter $\zeta$ for ensemble of RPSs with
  standard deviations $\sigma$ derived from the curve in Fig.~\ref{fig2}b
  (solid curve) and CPPSs (solid curve
  with $\ast$); $N_{L}=700$.}
\label{fig8}
\end{figure}

\begin{figure}  
 \includegraphics[scale=0.3]{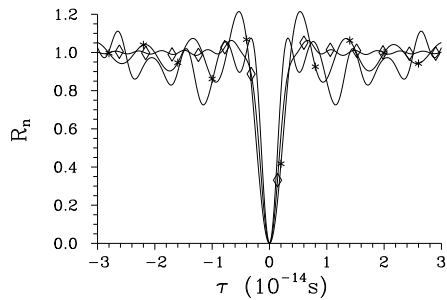}
 \caption{Coincidence-count rate $R_{n}$ as it depends on relative time
 delay $\tau$ for one realization of RPS (solid curve),
 CPPS (solid curve with $\ast$), and
 an ensemble of RPSs (solid curve with $\diamond$);
 $\sigma=2.1\times10^{-6}$m, $\zeta=2.5\times10^{6}{\rm m}^{-2}$, $N_{L}=700$.}
\label{fig9}
\end{figure}

A detailed analysis of the expression that gives the
coincidence-count rate $R_{n}$ in a Hong-Ou-Mandel interferometer
reveals an important property: the rate $R_{n}$ does not depend on
phase variations along the signal- and idler-field spectra in cw
regime. This property is frequently referred as nonlocal
dispersion cancellation \cite{Steinberg1992,PerinaJr1999a}. It
follows that entanglement time $\Delta\tau^{{\rm HOM}}$ is
inversely proportional to spectral widths $ \Delta S_{s} $ and $
\Delta S_{i} $ of the signal and idler fields despite their
complex profiles. We note that the entanglement time
$\Delta\tau^{{\rm HOM}}$ is determined by a temporal extension
(FWHM) of the coincidence-count interference pattern formed by the
rate $R_{n}(\tau)$. Entanglement time $\Delta\tau^{{\rm HOM}}$
thus shortens with increasing values of the standard deviation
$\sigma$ for RPSs. The dependence of entanglement time
$\Delta\tau^{{\rm HOM}}$ on the deviation $\sigma$ for an ensemble
of RPSs composed of 700 domains is shown in Fig.~\ref{8}. We can
see in Fig.~\ref{fig8} that entanglement times $\Delta\tau^{{\rm
HOM}}$ can be as short as several fs for sufficiently large values
of the deviation $\sigma$. This indicates that temporal quantum
correlations can be confined into an interval characterizing one
optical cycle provided that spectral phase variations in the
signal and idler fields are compensated. Entanglement times
$\Delta\tau^{{\rm HOM}}$ of CPPSs with the same signal-field
spectral widths $\Delta S_{s}$ are plotted in Fig.~\ref{fig8} for
comparison that reveals nearly identical entanglement times of
both types of structures.
Typical coincidence-count interference patterns given by $R_{n}$
for photon pairs generated in both types of structures are
compared in Fig.~\ref{fig9}. They demonstrate a close similarity
of photon-pairs behavior in a Hong-Ou-Mandel interferometer. There
occur typical oscillations at the shoulders of the interference
dips. Whereas regular oscillations characterize CPPSs, irregular
oscillations with larger amplitudes occur for individual
realizations of RPSs. However, widths of interference dips remain
practically unchanged for different realizations of RPSs.

\begin{figure}  
 \includegraphics[scale=0.3]{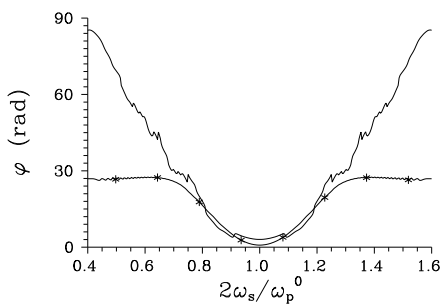}

 \caption{Phase $ \varphi $ of the two-photon spectral amplitude
  $ \Phi(\omega_s,\omega_p^0-\omega_s) $ as it depends on normalized signal-field frequency
  $ 2\omega_s/\omega_p^0 $ for one realization of RPS (solid curve) and CPPS
  (solid curve with $\ast$);
  $\sigma=2.1\times10^{-6}$m, $\zeta=2.5\times10^{6}{\rm m}^{-2}$, $N_{L}=700$.}
\label{fig10}
\end{figure}

\begin{figure}  
 (a) \includegraphics[scale=0.3]{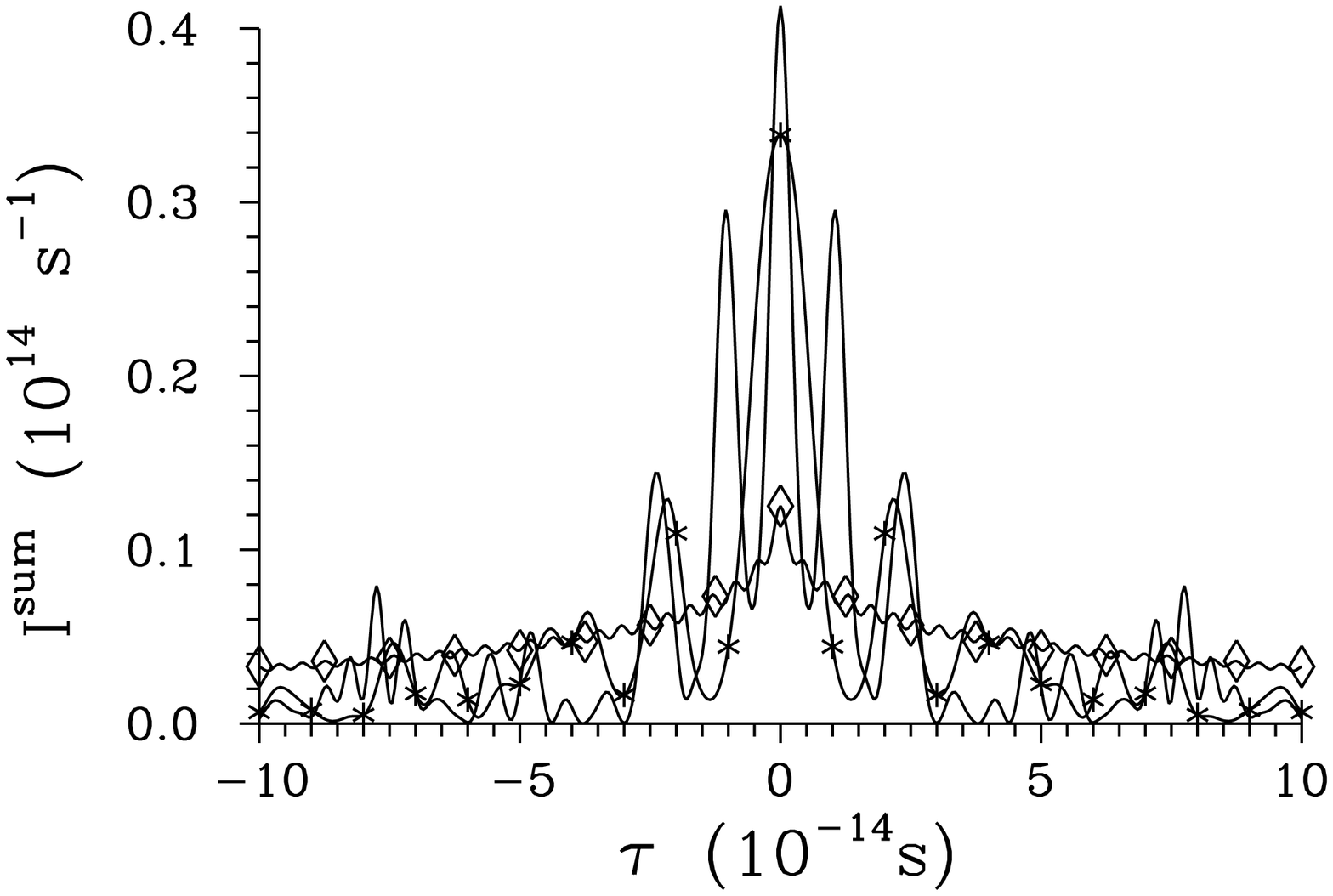} \vspace{7mm}

 (b) \includegraphics[scale=0.3]{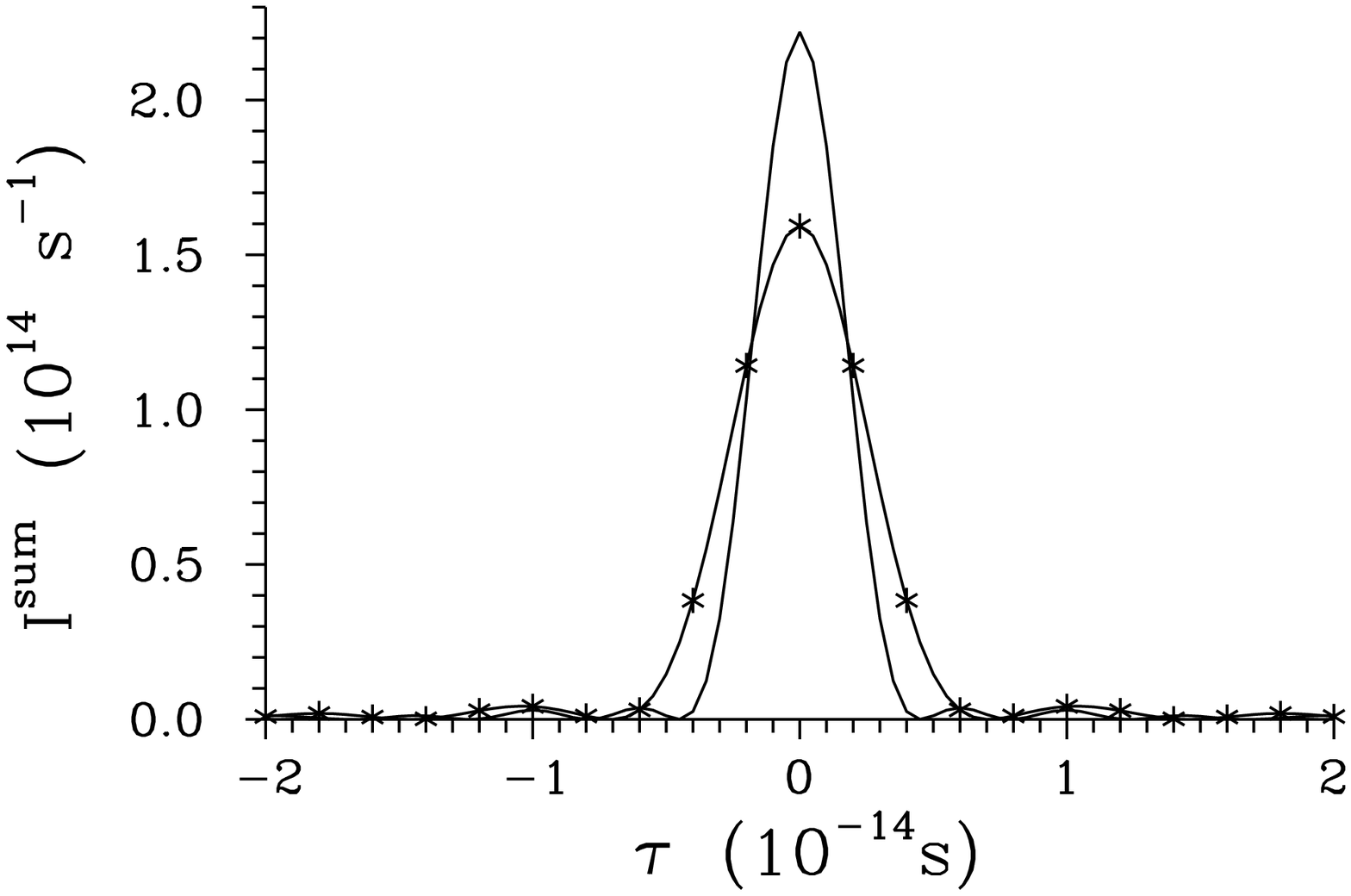}

 \caption{Sum-frequency field intensity $I^{{\rm sum}}$ as a function of
  relative time delay $\tau$ for one realization of RPS (solid curve), CPPS (solid curve with
  $\ast$), and an ensemble of RPSs (solid curve with $\diamond$).
  In (a) quadratic chirp in the signal-field amplitude spectrum is compensated
  for one realization of RPS and CPPS; in (b) complete spectral
  phase compensation is assumed. The curves are normalized such that
  $\int_{-\infty}^{\infty}d\tau I^{{\rm sum}}(\tau)=1$; $\sigma=2.1\times10^{-6}$m,
  $\zeta=2.5\times10^{6}{\rm m}^{-2}$, $N_{L}=700$.}
\label{fig11}
\end{figure}

Correlation times $\Delta\tau^{{\rm sum}}$ emerging from
sum-frequency generation are in general longer than entanglement
times $\Delta\tau^{{\rm HOM}}$ observed in a Hong-Ou-Mandel
interferometer because of a strong phase modulation along the wide
signal- and idler-field spectra $S_{s}$ and $S_{i}$ (see
Fig.~\ref{fig10}).
Correlation times $\Delta\tau^{{\rm sum}}$ can be even an order of
magnitude greater compared to entanglement times $\Delta\tau^{{\rm
HOM}}$ for structures with ultra-broadband spectra. However, phase
modulation along the spectrum can be compensated to certain extent
which gives shorter correlation times $\Delta\tau^{{\rm sum}}$.
CPPSs have more regular spectral phase behavior (as demonstrated
in Fig.~\ref{fig10}) and quadratic phase compensation is usually
sufficient to provide wave-packets several fs long. As for
individual realizations of RPSs, quadratic compensation is less
efficient because of more irregular phase spectral behavior.
Despite this values of temporal constants typical for chirped
structures can be approached [see Fig.~\ref{fig11}(a)]. Provided
that an ideal phase compensation is reached, both types of
structures give comparable results [see Fig.~\ref{fig11}(b)] and
are capable to generate photon pairs with wave-packets extending
over the duration of one optical cycle. Experimentally, pulse
shapers have been developed for this task and their capabilities
in the area of photon pairs have already been demonstrated
\cite{Dayan2007}. Comparison of the results obtained with
quadratic and ideal compensations reveals that correlation times
$\Delta\tau^{{\rm sum}}$ are approx. two times larger if we
restrict ourselves to quadratic compensation. Also the value of
quadratic chirp that needs compensation differs for individual
realizations of RPS. This requires an adaptive phase compensator.
On the other hand phase compensation in case of CPPS can be
reached in a simpler way, e.g., by inserting a peace of suitable
material of defined length \cite{Brida2009,Sensarn2010}. Despite
this RPSs are challenging both for basic physical experiments as
well as metrology applications.

\section{Correlations in the transverse plane}

In order to describe spatial properties of the signal and idler
beams (in the transverse plane) a simple generalization of the
model presented in Sec.~II has to be developed. The inclusion of
phase-matching conditions also in the directions along the $ x $
and $ y $ axes and assumption of spectrally-flat transverse
pump-beam profile $ E_{p\perp}(x,y) $ result in the following
separable form of a two-photon spectral amplitude $ \Phi $ that
additionally depends on radial ($ \vartheta_s $, $ \vartheta_i $)
and azimuthal ($ \varphi_s $, $ \varphi_i $) signal- and
idler-field emission angles (see Fig.~\ref{fig12}):
\begin{figure}   
 \includegraphics[scale=0.75]{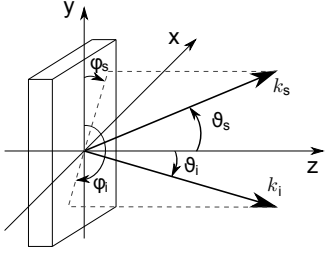}
 \caption{Geometric scheme for the description of
 spatial properties. Direction of the signal- (idler-)field wave vector $ k_s $ ($ k_i $)
 is given by radial $ \vartheta_s $ ($ \vartheta_i $) and azimuthal $ \varphi_s $
 ($ \varphi_i $) emission angles.}
\label{fig12}
\end{figure}
\begin{eqnarray}   
 \Phi(\omega_{s},\omega_{i},\vartheta_{s},\varphi_{s},\vartheta_{i},\varphi_{i}
  )&=& \Phi_z(\omega_{s},\omega_{i},\vartheta_{s},\varphi_{s},
  \vartheta_{i},\varphi_{i})  \nonumber \\
 & & \mbox{} \hspace{-15mm} \times \Phi_{xy}(\omega_{s},\omega_{i},
   \vartheta_{s},\varphi_{s},\vartheta_{i},\varphi_{i}),
\end{eqnarray}
where function $ \Phi_z $ arises from phase-matching conditions in
the $ z $ direction and function $ \Phi_{xy} $ originates in
phase-matching conditions in the transverse $ xy$ plane (see
Fig.~\ref{fig11}). Function $ \Phi_z $ can be derived in analogy
with the formula in Eq.~({5}):
\begin{eqnarray}   
 \Phi_z(\omega_{s},\omega_{i},\vartheta_{s},\varphi_{s},\vartheta_{i},\varphi_{i}) & = &
  g(\omega_{s},\omega_{i})E_{p}^{(+)}(\omega_{s}+\omega_{i}) \nonumber\\
 &  & \mbox{} \hspace{-25mm} \times
  F(\Delta k(\omega_{s},\omega_{i},\vartheta_{s},\varphi_{s},\vartheta_{i},\varphi_{i} )),
\label{36}
\end{eqnarray}
where the stochastic function $ F $ has been introduced in
Eq.~(\ref{6}). Phase-matching conditions in the $ xy$ plane give
the function $ \Phi_{xy} $ the following form:
\begin{eqnarray}   
 \Phi_{xy}(\omega_{s},\omega_{i},\vartheta_{s},\varphi_{s},\vartheta_{i},\varphi_{i})
  &=& \int_{-\infty}^{\infty}dx\int_{-\infty}^{\infty}dy E_{p\perp}(x,y)
  \nonumber \\
 & & \mbox{} \hspace{-15mm} \times \exp\left(i\Delta
  k_{x}x+i\Delta k_{y}y\right)
\label{37}
\end{eqnarray}
that includes a pump-beam amplitude profile $ E_{p\perp}(x,y) $ in
the transverse plane. Assuming normal incidence of the pump beam,
the cartesian components of nonlinear phase mismatch in
Eqs.~(\ref{36}) and (\ref{37}) can be written as:
\begin{eqnarray}   
 \Delta k_{x} &=&
  k_{s}(\omega_{s})\sin(\vartheta_{s})\sin(\varphi_{s})+k_{i}(\omega_{i})\sin(\vartheta_{
  i}) \sin(\varphi_{i})  \nonumber \\
  \Delta k_{y} &=& k_{s}(\omega_{s})\sin(\vartheta_{s})\cos(\varphi_{s})+k_{i}(\omega_{i})
   \sin(\vartheta_{i})\cos(\varphi_{i}) \nonumber \\
  \Delta k_{z} &=& k_{p}(\omega_{s}+\omega_{i})-k_{s}(\omega_{s})\cos(\vartheta_{s})
   -k_{i}(\omega_{i})\cos(\vartheta_{i}). \nonumber \\
  & &
\label{38}
\end{eqnarray}
We assume a Gaussian pump-beam transverse profile in numerical
calculations: $ E_{p\perp}(x,y)= 1/(\pi\Delta x_p \Delta y_p)
\exp[ - (x/\Delta x_p)^2 - (y/\Delta y_p)^2 ] $ and $ \Delta x_p $
($ \Delta y_p $) stands for the pump-beam width along the $ x $ ($
y $) direction.

\begin{figure}   
 (a)\includegraphics[scale=0.85]{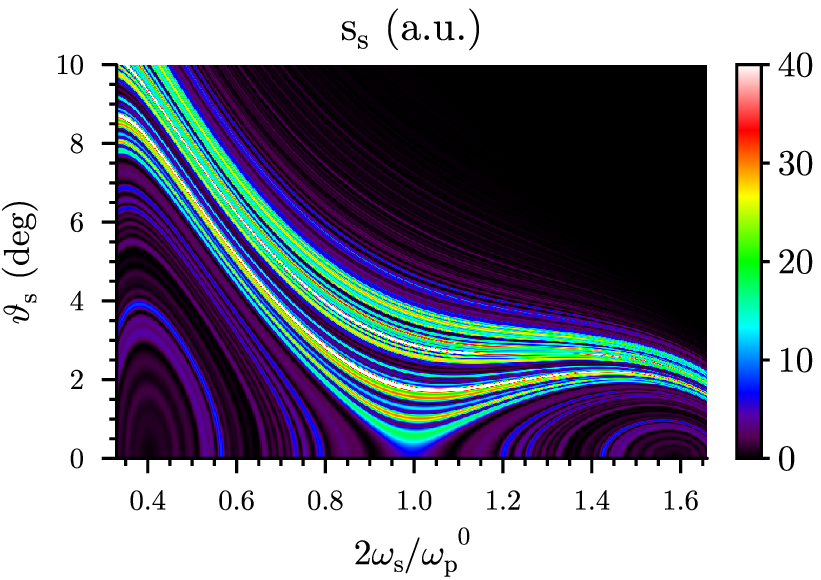}

 (b)\includegraphics[scale=0.85]{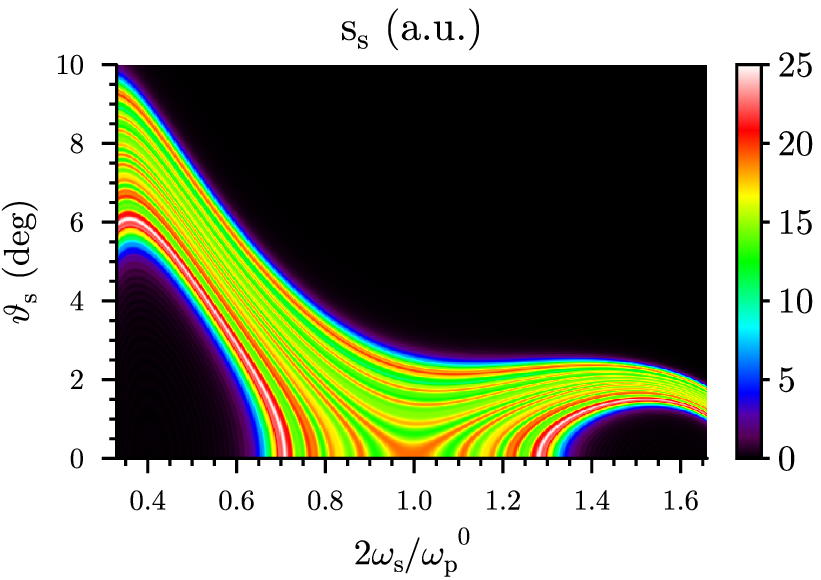}

 (c)\includegraphics[scale=0.85]{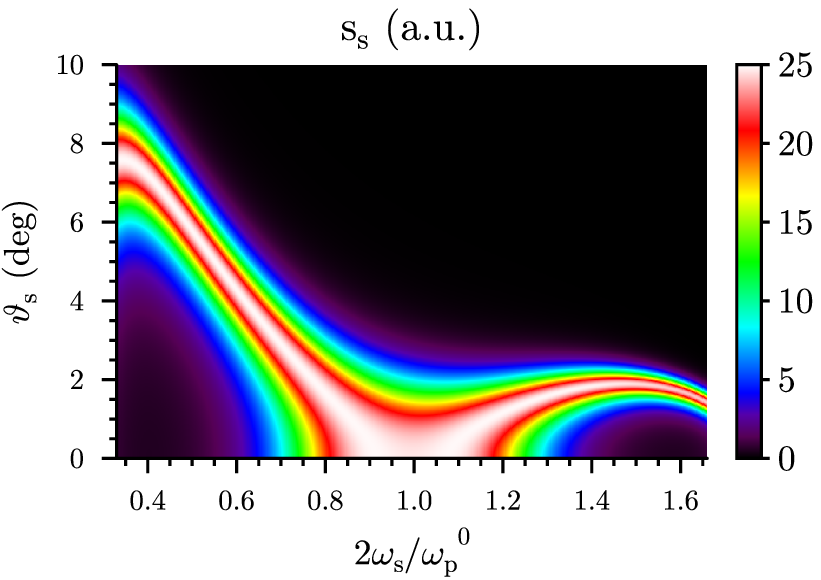}
 \caption{Map of signal-field spectral density $ s_s $ as it
   depends on signal-field radial emission angle $\vartheta_{s}$ for (a) one
   realization of RPS, (b) an ensemble of RPSs, and (c) CPPS; $ \varphi_s =
   0 $~deg, $\sigma=2.1\times 10^{-6} $m, $\zeta=2.5\times10^{6}$m$ {}^{-2}$,
   $N_{L}=700$.}
\label{fig13}
\end{figure}

We first pay attention to transverse properties of the signal beam
only. Its spectral density $ s_s $ defined as
\begin{eqnarray}   
 s_{s}(\omega_{s},\vartheta_{s},\varphi_{s}) &=& \sin(\vartheta_s) \int
  d\omega_{i}\int d\vartheta_i  \sin(\vartheta_i) \int d\varphi_i \nonumber \\
 & & |\Phi(\omega_{s},\omega_{i},
  \vartheta_{s},\varphi_{s},\vartheta_{i},\varphi_{i})|^{2}
\label{39}
\end{eqnarray}
depends on the signal-field radial ($ \vartheta_s $) and azimuthal
($ \varphi_s $) emission angles. As we study photon-pair emission
near the collinear geometry, the signal beam (as well as the idler
beam) has rotational symmetry along the $ z $ axis. The dependence
of spectral density $ s_s $ on signal-field radial emission angle
$ \vartheta_s $ is shown in Fig.~\ref{fig13}. Investigating one
realization of RPS we observe a typical 'strip-like' behavior
depicted in Fig.~\ref{fig13}(a). Fixing the value of radial
emission angle $ \vartheta_s $ spectrum $ s_s(\omega_s) $ is
composed of many peaks occurring at positions specific for the
studied realization [compare also with Fig.~\ref{fig4}(a)]. Each
peak changes continuously its central frequency as the radial
emission angle $ \vartheta_s $ moves. We note that this is typical
also for layered structures that form band-gaps
\cite{PerinaJr2006}. Averaging over many realizations of RPSs
smoothes this 'strip-like' behavior [see Fig.~\ref{fig13}(b)] and
leads to that resembling CPPSs [compare Figs.~\ref{fig13}(b) and
\ref{fig13}(c)]. In these cases spectral splitting is observed
\cite{Hamar2010}. This behavior originates in phase-matching
conditions along the $ z $ direction.

\begin{figure}    
 \includegraphics[scale=0.85]{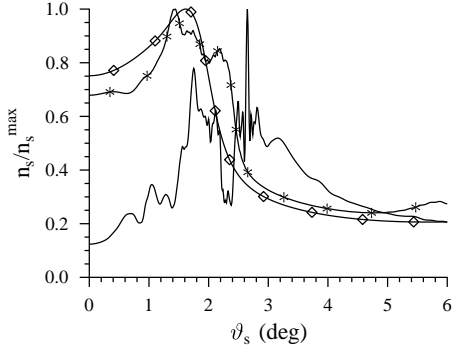}
 \caption{Profile of density $ n_s $ of signal-field photon numbers
  as a function of signal-field radial emission angle $ \vartheta_s $ for
  one realization of RPS (solid curve), CPPS (solid curve with $ \ast $), and
  an ensemble of RPSs (solid curve with $ \diamond $);
  $ n_s^{\rm max} = {\rm max}_{\vartheta_s}[n_s(\vartheta_s)] $.
  Plane-wave pumping is assumed; $ n_s(\vartheta_s,\psi_s) = \int d\omega_s
  s_s(\omega_s,\vartheta_s,\varphi_s) $; $ \varphi_s = 0 $~deg,
  $\sigma=2.1$~$\times 10^{-6} $m,
  $ \zeta=2.5\times 10^{6}$m$ {}^{-2}$, $N_{L}=700$.}
\label{fig14}
\end{figure}

Integration of spectral densities $ s_s $ over the signal-field
frequency $ \omega_s $ gives us densities $ n_s $ of photon-pair
numbers that are plotted in Fig.~\ref{fig14} for the structures
studied in Fig.~\ref{fig13}. Whereas the profile of density $
n_s(\vartheta_s) $ is complex for one realization of RPS, typical
shapes with one maximum around a nonzero value of $ \vartheta_s $
characterize the profiles of density $ n_s(\vartheta_s) $ for an
ensemble of RPSs and CPPS.

\begin{figure}  
 (a) \includegraphics[scale=0.85]{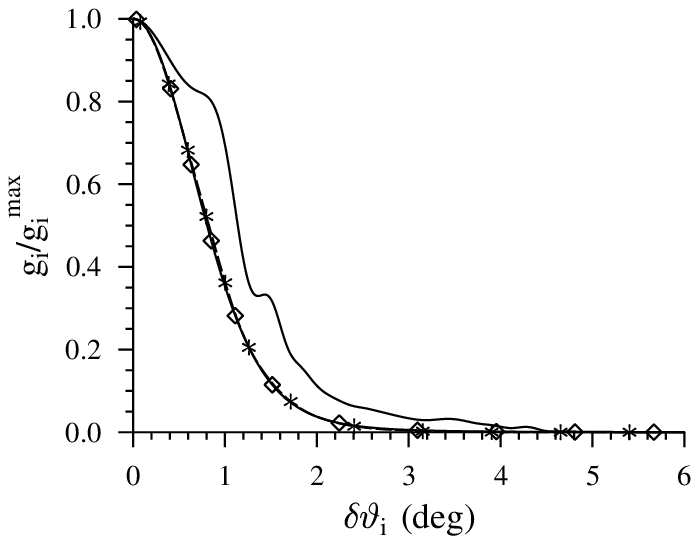}

 (b)\includegraphics[scale=0.85]{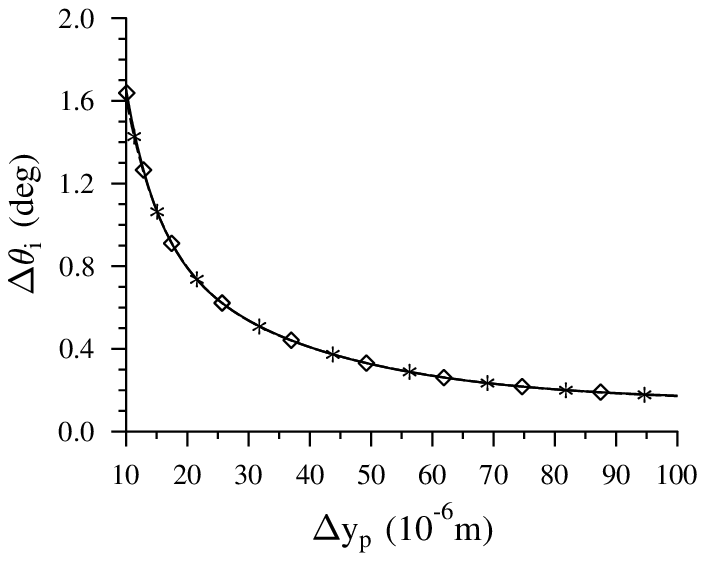}

 \caption{(a) Radial profile $ g_i(\vartheta_i) $ of the correlated area
 for $ \Delta y_p = 1 \times 10^{-5} $m
 and (b) radial width $ \Delta \vartheta_i $ of the correlated area as it
 depends on pump-beam width $ \Delta y_p $ for one realization of RPS (solid curve),
 CPPS (solid curve with $ \ast $), and an ensemble of RPSs (solid curve with $ \diamond $); $ \vartheta_i =
 \vartheta_i^0 + \delta \vartheta_i $; $ g_i^{\rm max} =
 {\rm max}_{\vartheta_i}[ g_i(\vartheta_i) ] $. Radially symmetric pump beam
 is assumed, i.e. $ \Delta x_p = \Delta y_p $;
 $ \varphi_s = \vartheta_s = 0 $~deg, $ \varphi_i = 180 $~deg, $ \vartheta_i^0 = 0 $~deg;
 $\sigma=2.1 \times 10^{-6} $m, $\zeta=2.5\times 10^{6}$m${}^{-2}$,
 $N_{L}=700$. }
\label{fig15}
\end{figure}

Correlated area $ g_i $ of an (idler) photon in a pair represents
spatial analogy to entanglement time and characterizes
correlations of photon twins in the transverse plane. By
definition, it gives probability of emitting an idler photon into
radial emission angle $ \vartheta_i $ and azimuthal emission angle
$ \varphi_i $ provided that its signal twin has been emitted in a
fixed radial emission angle $ \vartheta_s $ and azimuthal emission
angle $ \varphi_s $, i.e.:
\begin{eqnarray}   
 g_i(\vartheta_{i},\varphi_{i};\vartheta_{s},\varphi_{s}) &=& \sin(\vartheta_s)
  \sin(\vartheta_i) \nonumber
   \\
 & & \hspace{-2cm} \mbox{} \times \int d\omega_s  \int d\omega_i  |\Phi(\omega_{s},\omega_{i},\vartheta_{s},
  \varphi_{s},\vartheta_{i},\varphi_{i})|^2.
\label{40}
\end{eqnarray}
Because we mainly pay attention to beams propagating in the
vicinity of the $ z $ axis, we assume that the signal photon is
emitted along the $ z $ axis ($ \vartheta_{s} = \varphi_{s} = 0
$~deg). The correlated area as described by function $ g_i $ in
Eq.~(\ref{40}) then has rotational symmetry and its profiles along
the radial emission angle $ \vartheta_i $ for CPPS and an ensemble
of RPSs nearly coincide, as documented in Fig.~\ref{fig15}(a). On
the other hand, broader profiles are typical for individual
realizations of RPSs. These individual realizations form compact
correlated areas without large local peaks (compare with
Fig.~\ref{fig4} where spectrum $ S_s $ for one realization of RPS
is plotted). The width $ \Delta \vartheta_i $ of the correlated
area along the radial angle $ \vartheta_i $ depends in general on
phase-matching conditions along the $ z $ and $ \vartheta_i $
axes. Thus length of the structure, pump-field (temporal) spectral
width as well as width of the pump-beam waist determine together
the width $ \Delta \vartheta_i $ (for more details, see
\cite{Hamar2010}). For example, focusing the pump beam, values of
the radial width $ \Delta \vartheta_i $ can be varied nearly by
one order of magnitude [see Fig.~\ref{fig15}(b)]. This behavior
can be easily explained by the fact that the more the pump beam is
focused, the wider its spatial spectrum in the transverse plane,
and so the weaker the phase-matching conditions in this plane.

\section{The role of temperature}

\begin{figure}   
 \includegraphics[scale=0.85]{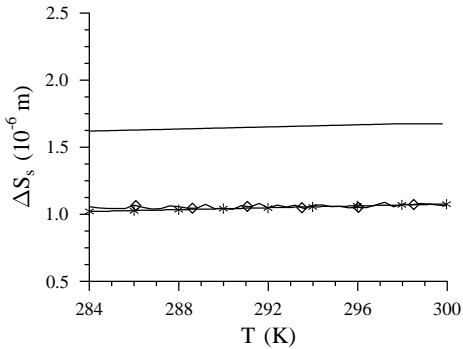}
 \caption{Signal-field spectral width $ \Delta S_s $ as it depends on
 temperature $ T $ for one realization of RPS (solid curve), CPPS
 (solid curve with $ \ast $), and an ensemble of RPSs (solid curve with $ \diamond $);
 $ \sigma = 2.1 \times 10^{-6} $m, $ \zeta=2.5 \times 10^{6} $m$ {}^{-2} $,
 $ N_{L}=700 $.}
\label{fig16}
\end{figure}

We have seen that an ensemble of RPSs and a CPPS have similar
properties. This close similarity is preserved also when studying
temperature dependencies \cite{Nasr2008} that are in general weak.
On the other hand, behavior of individual realizations of RPSs
manifests a stronger temperature dependence. However, influence of
temperature varies from realization to realization. Whereas
properties of the realization of RPS studied above do not
considerably change with temperature (see Fig.~\ref{fig16} for the
signal-field spectral width $ \Delta S_s $ in the temperature
range from $ 284 $ to $ 300 $~K), other realizations are more
prone to the change of temperature. This can be conveniently used
for efficient temperature modifications of properties of photon
pairs. We note that these effects have their origin in temperature
dependence of indexes of refraction \cite{Nasr2008}.

\section{The role of small random (fabrication) errors}

\begin{figure}  
 \includegraphics[scale=0.85]{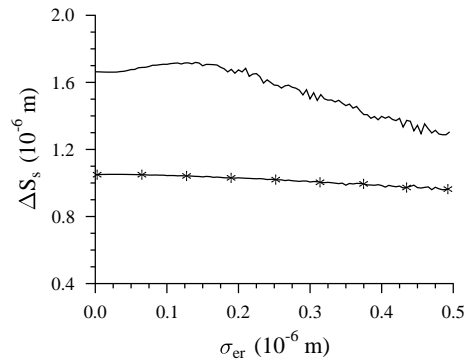}
 \caption{Signal-field spectral width $ \Delta S_s $ as a function of
 variance $ \sigma_{\rm er} $ of the fabrication error for
 one realization of RPS (solid curve) and CPPS (solid curve with $ \ast $).
 Averaging over the fabrication error was done in 1000
 randomly chosen positions;  $ \sigma = 2.1 \times 10^{-6} $m,
 $ \zeta=2.5 \times 10^{6} $m$ {}^{-2} $,
 $ N_{L}=700 $.}
\label{fig17}
\end{figure}

In the fabrication process, a small random error necessarily
occurs \cite{Fejer1992}. This error is sometimes called a duty
cycle error and, in general, leads to lowering of photon-pair
emission rates \cite{Pelc2010}. Considering spectral widths, they
are resistant against this error in uniformly periodically-poled
crystals \cite{Fejer1992} [see also Eq.~(\ref{18}) valid for
`weakly-random' structures]. On the other hand, spectral widths
are slightly reduced in CPPSs as documented in Fig.~\ref{fig17}.
We can see in Fig.~\ref{fig17} that a (large) fabrication error
with variance $ \sigma_{\rm er} = 5 \times 10^{-7} $m results in
the reduction of signal-field spectral width $ \Delta S_s $ only
by approx. 10~\%. Individual realizations of RPSs are much more
sensitive to the fabrication error. The observed spectral changes
depend on individual realizations. As an example, the signal-field
spectral width $ \Delta S_s $ of the sample analyzed above
decreases with the increasing variance $ \sigma_{\rm er} $ of the
fabrication error. This is natural, because spectrum of this
realization is broader compared to the ensemble mean value. We
note that it holds also here that the narrower the signal-field
spectrum, the greater the photon-pair generation rate $ N $ and
vice versa.

\section{The role of ordering in chirped periodically-poled structures}

\begin{figure}  
 \includegraphics[scale=0.85]{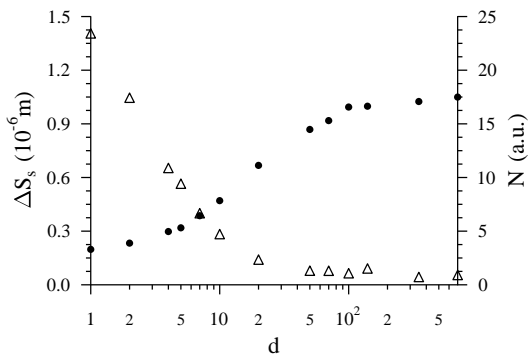}
 \caption{Signal-field spectral width $ \Delta S_s $ (solid curve with $ \bullet $) and
 photon-pair generation rate $ N $ (solid curve with $ \triangle $) as functions of segment length
 $ d $. Averaging over 1000 random positions of segments was
 used in calculations. $ \zeta=2.5 \times 10^{6} $m$ {}^{-2} $,
 $N_{L}=700$.}
\label{fig18}
\end{figure}

The benefit of ordering of individual domains by their lengths in
CPPS can be quantified as follows. We take an ordered structure
and divide it into segments containing $ d $ domains. We then
randomly position these segments in a new artificial structure and
finally obtain mean values of physical quantities after averaging
over random positions. In the limiting case of $ d=1 $ we have a
completely random structure similar to those studied above. It can
be shown that the signal-field spectral width $ \Delta S_s $
decreases with the decreasing segment length $ d $ (see
Fig.~\ref{fig18}). This is accompanied by an increase of
photon-pair generation rate $ N $. This behavior reflects the fact
that spectra of the fields coming from individual domains are
combined in a more constructive way in the central spectral area
with the increasing randomness (decreasing value of segment length
$ d $).
The graph in Fig.~\ref{fig18} also demonstrates that the
requirement for the same spectral widths $ \Delta S_s $ of RPSs
and CPPS inevitably implies that the histogram of domains' lengths
for RPSs is broader than that obtained for CPPS.

\section{Conclusions}

Properties of photon pairs generated in randomly poled structures
have been found quantitatively similar to those characterizing
chirped periodically-poled structures. Especially, ultra broadband
signal and idler fields can be emitted in randomly poled
structures. The accompanying sharp mutual temporal correlations of
the signal and idler fields can even reach the duration of a
single-photon cycle (several fs). Photon-pair generation rates
depending linearly on the number of domains are specific to random
structures. Stronger temperature dependencies of parameters
characterizing photon pairs in random structures (their individual
realizations) compared to those found in chirped
periodically-poled crystals have been observed. In general,
application potential of randomly poled structures similar to that
of chirped periodically-poled structures has been revealed.
Contrary to chirped periodically-poled structures randomly poled
structures do not require high precision in their fabrication.
This is a great promise for the use of randomly poled structures.

\acknowledgments Support by projects IAA100100713 of GA AV \v{C}R,
COST OC 09026, 1M06002 of the Czech Ministry of Education and the
Operational Program Research and Development for Innovations -
European Social Fund (CZ.1.05/2.1.00/03.0058) are acknowledged.
J.~S. was supported by the project PrF-2010-009 of Palack\'y
University.

\bibliographystyle{apsrev4-1}
\bibliography{Perina}

\end{document}